# GETTING READY FOR LISA

The Data, Support, and Preparation Needed to Maximize US Participation in Space-Based Gravitational Wave Science

A report of the
**NASA LISA STUDY TEAM**
Science Support Taskforce

J. Bellovary, P. Bender, E. Berti, W. Brown, R. Caldwell,
N. Cornish, J. Darling, M. Digman, M. Eracleous,
E. Ferrara, K. Gultekin, Z. Haiman, K. Holley-Bockelmann,
B. Kelly, J. Key, S. Larson, X. Liu, S. McWilliams,
P. Natarajan, D. Shoemaker, D. Shoemaker, K.L. Smith,
M. Soares-Santos, R. Stebbins

February 28, 2020

# TABLE OF CONTENTS

# EXECUTIVE SUMMARY

Dr. Paul Hertz, Director of Astrophysics at NASA Headquarters, has tasked the NASA LISA Study Team (NLST) to study how NASA might support US scientists to participate in the scientific exploitation of LISA data.  The LISA mission is led by ESA, and many of its Member States and their scientific communities will participate in all aspects of the mission.  The NLST is specifically charged to provide an analysis of how US scientists might participate in the science ground segment and access LISA data through public release.   The charge is broken into three tasks and 12 associated questions.

At the time of this report, the LISA Project at ESA is in the middle of Phase A, early in formulation.  The outlines of the ground segment, the nature of the data, and the details of the data analysis are still largely undefined.  In this document, the NLST has taken the current understanding of these components of the mission and extrapolated to a projected ground segment, data products and data analyses in order to represent what a future LISA user community might experience.  Of necessity, those projections are incomplete and speculative, and should not be construed as final or agreed upon by ESA, the LISA Consortium, or NASA.

With that caveat, we have arrived at the following general findings:
- A broad spectrum of US researchers are expected to want access to LISA data products, and based on our canvassing of the community, only half of those are willing to join a Consortium for access.

- Those researchers will carry out a broad spectrum of research, dependent on the extent of NASA's support.

- Pre-cursor science will prepare the community and enhance the research return. Notable examples are: study and simulation of data analysis methods; theoretical studies of astrophysical scenarios that could produce coincident gravitational-wave, electromagnetic, and neutrino observations; theoretical study and large scale simulations of source populations; numerical prediction of sufficiently accurate waveforms spanning the anticipated range of source parameters and physical processes; and electromagnetic observations of LISA source types.

- The novel nature of the instrument and the data will stimulate many research specialties and many research activities to seek access to deep levels of data, to L1 and below.  The most demanding science depends on deep access, from pre-launch testing to the earliest phases of the data processing.

- Prompt, public release of data products, with suitable support, will increase the scientific productivity of the mission through increased participation and diverse thinking.

- Coordination with electromagnetic observatories, particle detectors and ground-based gravitational-wave interferometers will be essential to the production and scientific utilization of LISA data products, most notably through low-latency alerts. LISA data products and LISA science can be significantly improved with information from other channels.

- A fuller array of tools, documentation and resources will expedite the research of US scientists.

- The US research community would benefit from maximal, expeditious access to LISA data products and from participation at all levels of data analysis -- from the earliest data products to catalogs.

- The most pressing activities are: studying and simulating the data analysis process, familiarizing and training non-GW researchers to work with LISA data, cultivating early-career researchers and precursor science, formulating the US LISA ground segment, and engaging in the deepest details of instrument design, analysis and testing to the greatest extent possible.

- The science productivity of a facility-class mission is greatly enhanced by a *full-featured science center and an open access data model*. As other major missions have demonstrated, a science center acts as both a locus and an amplifier of research innovation, data analysis, user support, user training and user interaction. In its most basic function, a US Science Center could facilitate entry into LISA science by hosting a Data Processing Center and a portal for the US community to access LISA data products. However, an enhanced LISA Science Center could: support one of the parallel independent processing pipelines required for data product validation; stimulate the high level of research on data analysis that LISA demands; support users unfamiliar with a novel observatory; facilitate astrophysics and fundamental research; provide an interface into the subtleties of the instrument to validate extraordinary discoveries; train new users; and expand the research community through guest investigator, postdoc and student programs. Establishing a US LISA Science Center *well before launch* can have a beneficial impact on the participation of the broader astronomical community by providing training, hosting topical workshops, disseminating mock catalogs, software pipelines, and documentation. Past experience indicates that successful science centers are established several years before launch; this early adoption model may be especially relevant for a novel mission like LISA.

# 1. BACKGROUND

In October 2019, Dr. Paul Hertz, Director of Astrophysics at NASA Headquarters, charged the augmented NASA LISA Study Team with three tasks to inform discussions with ESA regarding participation in the ground science segment and access to LISA data for the US community, through public release of data and data products. This report is the response to that charge. The charge, consisting of three tasks and 12 questions, is replicated in Appendix A for reference.

This report is organized into four sections. This section provides background on the LISA mission pertinent to the charge, as a reminder to the reader of the relevant mission characteristics that are unique to LISA. It specifically describes the characteristics of the data (§1.1), general content and flow of the data (§§1.2,1.3), data access points (§1.4), LISA science(§1.5), community surveys (§1.6), and applicable documents (§1.7).

Note that the LISA Project is in the middle of Phase A at this writing, and this report is, of necessity, based on an in-progress formulation of the mission. The applicable documents and the space and ground segment designs will certainly change as the formulation and development progress.

§§ 2, 3 and 4 each address one of the three tasks in the charge, and the subsections respond to associated questions (Appendix A).

The appendices contain a copy of the charge, the LISA Science Objectives and Investigations, the results from the LISA Interest Survey, detailed latency considerations and an extended list of anticipated data products.

## 1.1 OVERVIEW OF LISA DATA CHARACTERISTICS

At this time, early 2020, very few researchers have had any experience with GW data. LIGO-Virgo data are the only example with confirmed detections, and their analysis has been conducted almost wholly within the LIGO and Virgo Collaborations. As detailed in Table 1 of the previous NLST report (Ref. 4 in §1.8), LISA data will be different in many significant ways. This novelty is a critical consideration for the LISA ground segment, and for the researchers who will use the data and data products. This section reviews the special characteristics of LISA sources that will impact data, data analysis and derived science products.

The LISA detector senses the entire sky all the time. The signals from all perceivable sources are present in all the data. This will amount to tens of thousands of individual sources of many types with widely varying characteristics. There will also be a confusion continuum of unresolved galactic binaries, and possibly an astrophysical and/or cosmological background. There will be merger events weekly or more often, which can be used to trigger alerts *in advance* for other observing assets to follow up.

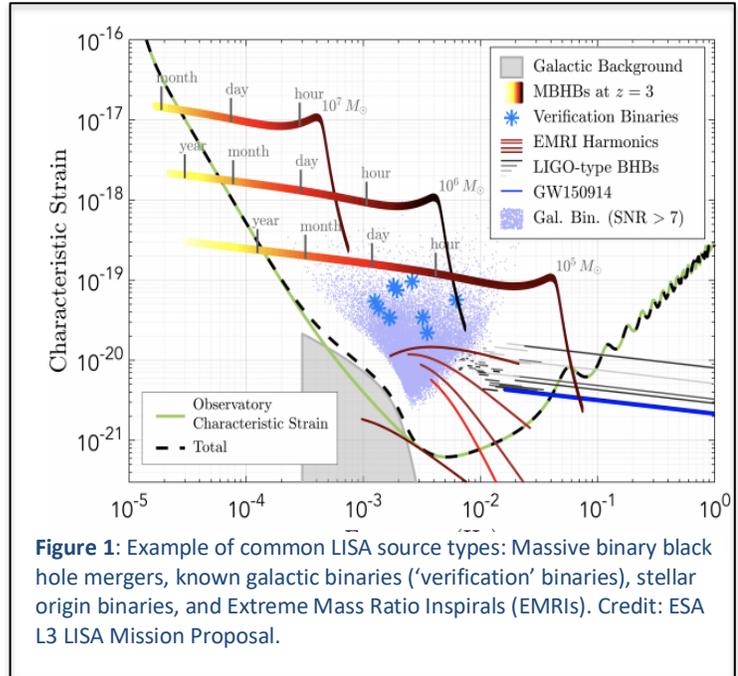

**Figure 1**: Example of common LISA source types: Massive binary black hole mergers, known galactic binaries ('verification' binaries), stellar origin binaries, and Extreme Mass Ratio Inspirals (EMRIs). Credit: ESA L3 LISA Mission Proposal.

In another contrast to the data from ground-based GW detectors, the LISA data stream will be dominated by thousands of compact binaries producing continuous signals, the frequency of which change very little during the life of the mission, but whose frequency, amplitude and phase are modulated by the orbital motion of the LISA detector constellation. As can be seen in Figure 1, taken from the ESA L3 Mission Proposal, however, there will also be:

- Compact binaries with intrinsically changing frequencies, such as mass-exchange binaries,
- Massive black hole binaries that merge and ring down after persisting in the data for days to years,
- Binaries that emerge from the confusion threshold as the length of the data set improves the frequency resolution, typically white dwarf binary systems,
- Wider binaries that evolve into the sensitivity band as their frequency chirps towards merger,
- More compact and lower-mass binaries that evolve out of the sensitivity band as they chirp towards merger at frequencies above the LISA sensitivity band, say, in the LIGO/Virgo band,
- Changing confusion limits, as the sensitivity pattern of the antenna moves across the Milky Way,
- And possibly stochastic backgrounds of uncertain character and/or unknown origin.

Identifying the individual astrophysical signals and separating them from the instrumental effects constitutes the bulk of the analysis effort to produce derived science products, such as source catalogs. In principle, a user with access to the GW strain data has access to all of the primary science data. In practice, a deep knowledge of the instrument, a vast command of astrophysics, and significant expertise in data analysis is essential to extract scientific results.

## 1.2 DATA TYPES

Several characteristics of the LISA data are unique and thereby affect the science ground segment and data products. Those characteristics are recounted here to help the reader understand the remainder of this report.

The LISA scientific instrument comprises 3 spacecraft and simultaneous data streams from many instrument channels representing the perceived lengths of each arm. Data from all three spacecraft are downloaded routinely to the Mission Operations Center. Several complicated processing steps are required to produce the lowest level data product usable by any researcher not intimately familiar with the design and construction of the spacecraft. Specifically, the primitive phasemeter data has to be converted to armlength changes and combined to produce strain data. Before that step the science data are swamped by laser frequency noise by many orders of magnitude. As with several other steps in the LISA data processing, the early critical steps are likely to be iterative.

Because the precise data analysis pipelines are still being developed, we caution that the data product descriptions here are likely to change and are not complete. For example, decisions will need to be made to determine what data quality or level will suffice for an alert to EM/Particle/Ground-based GW observers, and there is a debate as to what data products are considered strictly 'housekeeping' versus 'science'. We base our broad summary on current data flow and definitions of data levels as set out in the first draft of the Science Operations Assumptions Document (SOAD, ref. 2 in §1.8) wherein the basic data types are enumerated by the ESA LISA Project. The SOAD establishes the basic data products that ESA expects to produce, as of the Mission Consolidation Review (Nov. 2019).

**Initial data levels consist of both raw and pre-flight data products.**

**Pre-flight (PF)** spacecraft data: These entail a complete description of all components in each spacecraft, including pre-flight measurements and test results, mass, thermal, and control systems models, as well as information from industrial partners about performance, such as hardware test results, tolerances, traceable calibration, and unit conversions to physical units. Experiences in space missions such as Kepler[1] and WFIRST have emphasized the importance of both summary hardware data and large volumes of raw hardware test data to enable the deep dives needed to solve puzzles seen in the science data.

---

[1] https://trs.jpl.nasa.gov/bitstream/handle/2014/44744/10-2955_A1b.pdf?sequence=1

**Augmented Telemetry (ATM)** data: Data products in this category come from all payloads and systems in the constellation as they arrive into the Mission Operations Center (MOC). These data products have not been processed; they will contain errors and gaps and will arrive out of order. Many quantities, such as voltage or temperature, will be in non-physical units. ATM data include: laser power, star-tracker data, clock times, magnetometer measurements, temperature measurements, phasemeter measurements, etc. At the MOC, these data are combined in 'preliminary processing' with orbit data and ground station information in a rigorous step that involves continuous monitoring of the instrument and ground station performance.

## Preliminary processing yields low-level data products.

**Level 0 (L0)** data: Here, all raw data (including housekeeping) are converted to physical units, corrupted data are flagged and removed, and the data are time-tagged and ordered to a common epoch for housekeeping and in preparation for creating time-series strain data. Hence, data products in the L0 level can be thought of as the cleaned, calibrated, and time-ordered ATM data products.

**Level 1 (L1)** data: Noise from various instrument subsystems across the constellation are canceled in post-processing, creating *time series data products with extractable gravitational wave signals*. So-called Time-Delay Interferometry, or TDI, is a key algorithm in this process. L1 requires careful modeling and cancellation of time-varying instrument behavior, such as how changes in angle of the test-mass and spacecraft couple to the effective optical pathlength (tilt-to-length coupling, or TTL). L1 data are created with the lowest possible latency to serve in *'quick look'* transient alerts, though later and more careful reanalyses are expected as more is learned about the instrument and station-keeping performance in orbit.

## Extensive data analysis yields high-level data products:

The LISA mission proposal suggests that high-level data products are created within data processing centers (DPC), with close connection and iteration between DPCs and the SOC.

**Level 2 (L2)** data: Since there will be millions of GW sources all present in the data stream at once, the current plan to analyze the data is to construct solutions from the global and time-evolving L1 data stream. We envision at least two categories of L2 data: data products from the output of one or more *'quick look'* pipelines to capture and characterize high signal-to-noise point sources to alert EM/Particle/Ground-based GW observers, and an evolving *'global fit'* solution to the L1 data stream, based on priors from theoretical models, observational data, and computational waveforms of individual sources. The global fit will include both resolved and unresolved sources. Both categories will contain residuals from the analysis that are considered a valuable data product used to assess the health of the instrument and/or search for unmodeled signals, as are the priors assumed for processing. We emphasize that this global fit will likely be an iterative process that will include revisiting lower level data, using different waveform template suites and fitting some of the instrumental unknowns. This iteration will be necessary for a given body of data, and also as new data are collected.

**Level 3 (L3)** data: The main L3 data product is expected to be a catalog consisting of high-confidence resolved sources, complete with their measured physical parameters and uncertainties, such as mass, spin, eccentricity, measured sky position and distance, and suspected source type, e.g. 'supermassive black hole merger'. Also included in such a catalog would be various flags describing data quality, data reduction choices such as the class of waveform templates or stellar population model assumed, and caveats such potentially blended or disappearing sources. In addition, this catalog would provide posterior distributions of the measurements, time of event if transient, and a measure of confidence.

**Auxiliary Data Products (AUX)**: Material such as the algorithms, software, documented processing history, intermediate data products resulting from iterative processing steps, waveform templates, *residuals*, instrument model assumptions (such as TTL coupling model), astrophysical model assumptions (such as binary mass loss during the common envelope phase), etc. are also important data products *that span many levels.* For example, the software and selected intermediate data products for processing L1 from L0 are anticipated to be useful in tracking down inconsistencies in the higher-level data streams or to improve instrument sensitivity. Any analysis indicating e.g. a shortcoming of general relativity would require careful study of AUX products as part of gaining confidence in a conclusion. Key high-level AUX products may include data streams cleaned of bright sources, enabling dedicated reanalysis of individual loud events or searches for other, unanticipated, source types.

**Added-Value Data Products (AV)**: One of the most common classes of AV include L2 or L3 data cross-matched with other astrophysical data, such as a galaxy catalog. These AV may be used in post-processing, though they may also be useful as input into the L2 analysis.

---------------------------------------------------------------------------------------------------------------------------------

ESA is currently anticipating making L3 available through a public archive, the timing of which is under discussion. At present, there is discussion of when and how upstream products (e.g., those at L1, L2) will be made available publicly, and with what documentation. A NASA LISA Science Center might reasonably expect to host a mirror of these data archives for access by US researchers. Modalities of access to data before Level 1 is also under discussion at this time.

Two important features of LISA operations and data processing anticipated in the SOAD are **protected periods** and **alerts**. Protected periods are times when, e.g., a massive black hole binary merger event is predicted. Scheduled spacecraft downtime is postponed, data may be downlinked more frequently, and enhanced data processing could take place. The goal is the production of rapidly improving sky coordinates for the source with very low latency. Likewise, alerts of newly discovered sources and soon-to-merge sources will also be produced with the lowest possible latency to enable EM observatories and particle detectors to make coordinated observations of GW sources. This requires a low-latency analysis pipeline and rapid release mechanism, which is currently tied in the SOAD to L2, though this may change.

## 1.3 DATA FLOW

The SOAD describes basic assumptions about data flow between the Mission Operations Center (MOC), the Science Operations Center (SOC) and the final public archive at the ESA Science Data Center of the ESA Space Astronomy Centre. That flow is encapsulated in Figure 2, copied from the SOAD. This is meant to be illustrative and may not represent the final process.

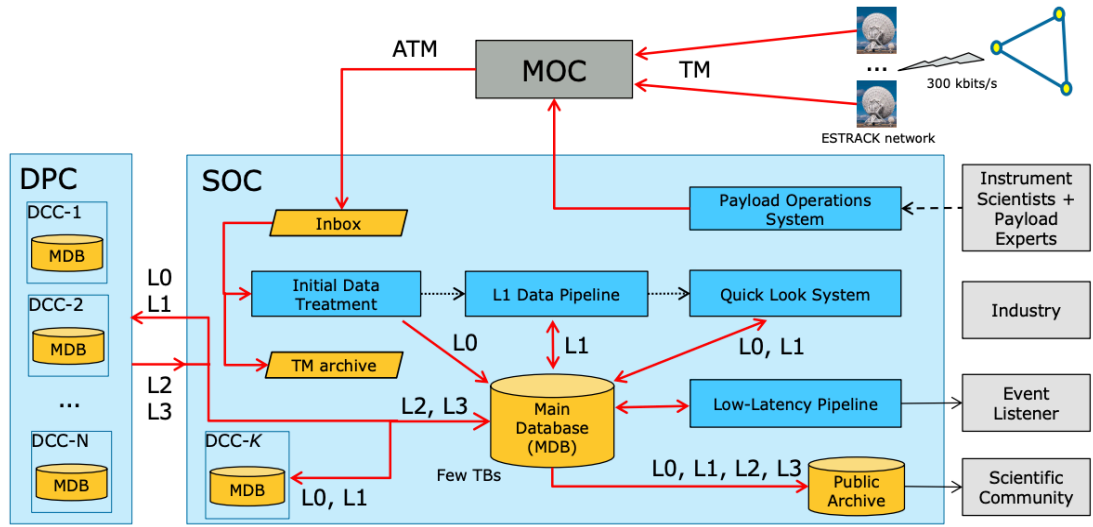

**Figure 2:** Overview of LISA data flow, from SOAD (rev. 2.3). ATM data flows from the MOC to the SOC where L0 and L1 data are produced. L0 and L1 data flow to the Distributed Data Processing Centers, operated by LISA Consortium members, to generate L2 and L3, which are returned to the SOC for archiving. L0-L3 will be made available to the scientific community through a public archive– discussions are underway to determine when, how, and with what level of documentation and support they are to be released.

The MOC sends daily ATM data to the SOC where L0 and L1 data are created. L0 and L1 data are sent to Distributed Data Processing Centers (DDPCs), operated by LISA Consortium members, to generate L2 and L3. The L0-L3 data are returned to the SOC for archiving. Although lower-level data products are intended to be released – discussions center around when, how, how often, and with what level of documentation and user support are they to be released to the public.

A quote from the SOAD illustrates the issues at hand: "The guiding principle for publication of the data products is to enable the scientific community to re-do any analysis of the data at least from L1 (TBC) to the L3 products. This requires that, in addition to the publication of the data itself, all algorithms (TDI, searches), software, and models (e.g. signal waveforms, dynamical GRS model (TBC if feasible)) used in the processing be made available. In addition, the processing history for any published data needs to be traceable and accessible. All data products will be public and all relevant software placed under an open source license." Specifically, there is a very active discussion about the data flow steps between L0 to L3 within ESA's Science Study Team (SST), the LISA Consortium, and the NLST. Generally, it is expected that the processing steps from L0 to L1 to L2 to L3 will be the subject of research and evolution up to and throughout the mission as the understanding of the instrument progresses and actual data become available.

## 1.4 Data Access

Various groups of technical professionals will have access to the data at different points during the data reduction process. The assumptions of the SOAD, as illustrated in Figure 2, give a notional picture of how different groups could access data products to carry out scientific investigations. Roughly, those groups and their access points are:

- MOC Personnel handle TM data from the downlink stations until it is passed to the SOC as ATM data. No scientific research opportunities are present here. All MOC personnel will be ESA engineers and technicians.

- SOC Personnel accept and archive ATM data, produce L1 data, forward it to the DDPCs, and receive and archive the L2 and L3 products. Hence, they will have access to all data products. The software for producing L1 data will be produced by instrument experts in the LISA Consortium in collaboration with SOC Personnel. There will likely be opportunity for scientific research and publication in situations where interpretation is dependent on the details of instrument design, construction and performance. Examples might be tests of General Relativity, detection of exotic sources, and unexpected discoveries. One might reasonably anticipate that publications on these topics could include authors from the SOC, the Consortium and the external scientific community. SOC Personnel will generally have special knowledge and insight into the LISA observatory and data processing, and will staff ESA's user support function. In their ESA user support role, SOC personnel may collaborate with users on scientific research, or may function as a helpdesk line to forward questions to relevant experts; this depends on the user support model ESA decides to adopt.

- The LISA Consortium will operate the DDPCs where L2 and L3 data are generated. Members of the Consortium will likely pursue many aspects of the science objectives and investigations; they come from all of the participating European member states, the US, and other countries, and are supported by their respective national agencies. They are also responsible for designing, developing, and maintaining the processing steps that produce L2 and L3, and can be expected to be intimately familiar with the process and products. This will be a major research effort from Phase A until the final data release in Phase F.

- External US researchers will have whatever public access modality granted by ESA. We expect that this policy will be negotiated with NASA. The time-varying nature of LISA data and the demand for timely alerts will necessitate multiple data releases. The number, latency, frequency and contents of those data releases are not yet decided. Public access via the internet ensures that the data archive offered by ESA, and likely mirrored by NASA and some DDPCs, can be accessed by researchers worldwide.

- Beyond mere access, conducting science investigations with the data will require extensive documentation and user support. Discussion is underway as to the timing and level of documentation and support provided, both within the LISA Consortium and to the public. We discuss user support in §4.1.

For the purposes of responding to the charge from Headquarters, a much more expansive discussion of data access is given in §3 on Value and Utility of LISA Data Products.

# 1.5 LISA Science

The baseline LISA science was established in the LISA Consortium's response to the ESA call for L3 mission concepts [Ref. 3 in §1.8], and 8 formal Science Objectives are captured in the LISA Science Requirements Document [Ref. 1 in §1.8], see Appendix C. However, the original proposal and the Science Requirements Document focus on science that can be done, with high certainty, with LISA data. *Interestingly, these objectives are not necessarily what the LISA Consortium can or will address; this is an ongoing discussion and is dependent on funding.*

Our expectation is that LISA will stimulate much broader scientific research, and NASA's science program will have to respond if it wishes to enable broader discovery. The following list is a high-level outline that is necessarily incomplete:

- **Instrument science** – GW detection is an extremely demanding measurement challenge, and research on understanding and characterizing the performance of LISA will certainly continue until the end of the project. Following the example of LISA Pathfinder, much of which will be reprised in science commissioning, the Consortium's Instrument Group, and probably select external experts will undoubtedly study the dominant noise processes in the instrument, both for the purpose of other data products, but also to understand the measurement technology and for the benefit of future GW missions.
- **Data analysis** – The LISA data analyses developed before launch will certainly evolve when confronted by real data. We expect that the early data analysis will spawn research efforts to find better correction algorithms, better source detection algorithms, better source waveforms, better alerts, better catalogs, etc. that will continue throughout the life of the project. These efforts may require producing new data, data levels, or data products. It is absolutely critical to anticipate and provide for research on the data analysis.
- **Observational astrophysics** – LISA GW data will contain observational information about compact objects in binary star systems, including white dwarfs, neutron stars, and black holes from stellar to supermassive sizes. LISA will be able to observe sources from the Milky Way to redshifts far beyond those reachable by any electromagnetic observations. These observations and coordinated EM observations, where achievable, will prompt research activities in: the population, formation and evolution of compact binaries, including mass exchange and tidal interactions; the structure of the Milky Way as mapped by compact binaries; the origin and evolution of stellar-mass black holes now being reported weekly by LIGO and Virgo, especially by multi-band GW observations by LISA and ground-based GW observatories; the existence and demographics of intermediate-mass black holes (IMBHs, $10^3$-$10^4$ $M_\odot$), for example in globular clusters and dwarf satellite galaxies; the seeds, evolution, merger history, mass function, spins, and kicks of massive black holes; the merger history of galaxies; the environment and dynamics in galactic nuclei close to the central engine; and time-domain astronomy enabled by LISA's advance alerts.
- **Computational and theoretical astrophysics** – A wide range of numerical and theoretical models will both support and be affected by LISA observations. One class science in this category involves generating waveforms to develop a deeper understanding of GW sources and astrophysical scenarios where extreme gravity dominates. For example, stellar evolutionary models will inform the waveform templates used to identify sources, and will be tested in turn by the analysis pipeline. Because of its inherent sensitivity (e.g., waveforms accurate to a fraction of a cycle over the hundreds of thousands of cycles spanned by the mission lifetime), LISA source detection needs waveforms that include all relevant physics. Another class of science involves the demographics of LISA observations to understand the interplay between GW sources and their hosts. For example, galaxy models, especially for the Milky Way and its satellites, will both drive the data analysis, and be tested by it. The very largest cosmological simulations are providing the best estimates of massive black hole populations, but they are still imperfect because of the unknown physics of black hole formation, feedback and evolution. Again, these computational models will both inform the LISA data analysis and will be tested by the sources produced.

- **Relativity and fundamental physics** – LISA observations will explore the nature of black holes and gravitation at a level not previously possible. The highest signal-to-noise ratio detections will make precision measurements possible that will eclipse those expected from ground-based detectors, including from those in operation during the time of the LISA mission. This will trigger research activity on: the properties of GWs; predictions for sources, especially the waveforms produced by numerical relativity and approximation methods; tests of General Relativity (GR) in the strong dynamical limit; exploration of new theoretical ideas beyond GR; and exotic sources such as cosmic strings, and new physics.
- **Observational and theoretical cosmology** – GWs constitute an entirely new method to determine a distance scale, one that is largely model-free, and hence has different systematics than the current methods which are in tension with each other. The expansion rate, the Hubble constant, other cosmological parameters, the earliest seed black holes, stochastic backgrounds, and exotic ingredients like cosmic strings can all be investigated with LISA data. New information on these topics will inform observational cosmology and benefit from and stimulate theoretical cosmology.

Two other aspects of LISA operations will have substantial impact on research activities. First, as with LIGO and Virgo, the LISA SOC will be issuing alerts for electromagnetic observatories and particle detectors to follow-up. The LIGO/Virgo experience to date suggests that this will engage an important fraction of ground and space astronomy assets. Over the next 15 years, the LIGO/Virgo experience may become less relevant, as other observing channels more efficiently respond to alerts. However, the considerably higher number and greater variety of LISA sources (including those that are extremely long-lived) will have a greater impact on the operations of contemporaneous observing assets. Whatever the balance, science ground segment planning from now forward should take serious consideration of coordination with the contemporaneous observatories and detectors.

Secondly, LISA is already a broadly international undertaking. ESA, of course, leads LISA, but the participating Member States and NASA will lead the science activities based on LISA data. Further, scientists worldwide is expected to have access to the science products through the ESA public archive, and mirrors hosted by Member States and/or NASA. Other nations (e.g., Japan) may yet join the mission. Consideration of the international context is critical in planning science activities at NASA.

## 1.6 LISA INTEREST SURVEY AND OTHER INPUT

For the purposes of this report, the NLST canvassed the research community about interest in LISA and desired support from NASA. A formal web-based survey was conducted, and informal input was sought from other sources. The content, distribution, and results of the LISA Interest Survey are reported in Appendix B, but we summarize it briefly here:

We drafted a 16-question survey to gauge U.S. researchers' needs and interest in using future LISA data. We announced the survey in the American Astronomical Society's monthly Newsletter, the Astronomers Facebook page, and on large (LSST, SDSS, PCOS) email listservers. We ran the survey for 7 weeks and received 193 responses.

Four questions were demographic. We had a diversity of respondents from over 100 unique institutions, a broad range of career stages, and a good mix of theorists and observers. About half of respondents were familiar with LIGO/Virgo alerts, however only a subset of respondents were familiar with LIGO/Virgo data in depth.

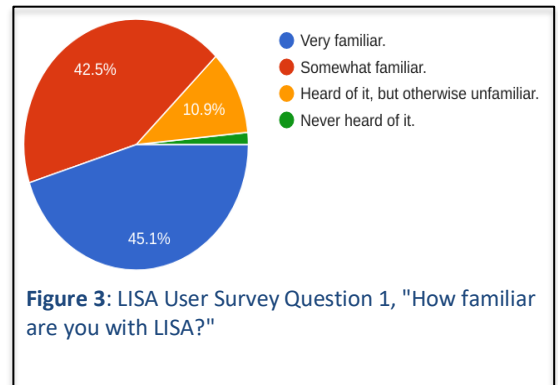

When asked "How familiar are you with LISA?", approximately half of the respondents were "very familiar" with LISA, half were not. Since the goal was to survey the broader community, we examined the results split by this familiarity response. Most results were common to both halves; the biggest difference was interest in low-level data (those familiar with LISA expressed more interest in low-level data).

**Figure 3**: LISA User Survey Question 1, "How familiar are you with LISA?"

The LISA-related questions typically asked respondents to rank-order a list of choices. For LISA data products, the highest-ranked choice was the catalog of resolved sources, followed by LISA alerts. For tools, the highest-ranked choice was a cross-matched LISA source catalog, followed by high-level analysis tools, and then recipes for common use cases

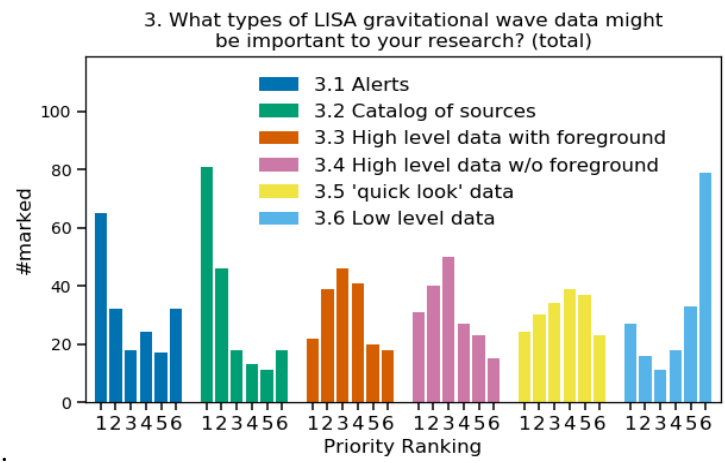

**Figure 4**: Total response to LISA User Survey Question 3, "What types of LISA gravitational wave data might be important to your research?". Note that roughly 73% of users rate alerts and highly-processed catalogs as their top priority.

Respondents overwhelmingly ranked written documentation as the top data-training choice, and ranked having a "general introduction" to LISA higher than other options. When asked "Would you be interested in exploring mock data now?", about a third said "yes," a third "maybe," and a third "no." The majority of respondents thought LISA data-training should happen close to launch.

Interestingly, we asked about LISA data products twice, the second time framing the question as "Would you be willing to join a large collaboration to use the following LISA data?" Responses are correlated with the question about interest level, but *half of respondents are unwilling to join a large collaboration to use LISA data.*

We had two write-in questions, asking what precursor science should be done, and asking for final thoughts. We received dozens of responses. For example, statements regarding data access and community support had the following flavor (these are direct quotes):

- Looking at the records of past facilities, the biggest impact is had when a large segment of the science community is involved; and the latter happens when high-level data products are provided, rather than requiring outsiders to acquire black belts in analysis (although of course a few would prefer the latter kind of access). Widespread uptake of open surveys, catalogs, *digested* information, available to the broadest community (and not just the constituent community) - this is what leads to high impact on science.

- LISA will be transformative – in the science, in the data, and in the mission itself. With such a huge potential for a paradigm shift, it's crucial to try to futureproof the data analysis with extensive documentation, lots of data from intermediate steps in the analysis, and archived information all the way down to the nuts and bolts of the spacecraft to triple-check amazing claims (think the Opera experiment and tachyons that were really some fiber optic glitch in the clock). One aspect of futureproofing is getting the broadest community involved to look at the data and the science in different ways. This could start now with various types of mock data and training.

- Don't make the same mistake as LIGO; release the data as quickly as possible so that the wider community can interface with it easily.

- In order for current graduate students to make contributions to LISA science, it is important that the pipeline for getting those graduate students into postdocs and ultimately faculty or permanent research positions is built now, so that  researches with an extensive knowledge base can be prepared to make the best possible use of LISA data, and mentor their own graduate students and postdocs when LISA launches in the 2030s. Financial and community support for precursor science in the 2020s, including source modelling, direct LISA modelling, and electromagnetic surveys to find and study LISA's galactic sources ahead of time are vital to ensuring there will be an active and experienced pool of scientists to promote LISA's science output.

- Make the data as open as possible. Produce collaboration-wide papers describing tools -- these can then be cited by all users of the data so that credit is given.

In addition, ground-based GW observatories are taking data in the LIGO-Virgo O3 run until end-April 2020, and alerts are being distributed to, and utilized by, the interested research community through NASA's GCN mechanism. Those alerts and responses to them are accessible through two mobile device applications, *GW Events* and *Chirp*. The NLST contacted the authors of those apps and solicited input from users to get a sense of how GW data is being used right now. The users of those apps comprise a large and active population of GW data consumers, including both a wide range of observational astronomers and a large population of amateur astronomers and enthusiastic citizens.

## 1.7 APPLICABLE DOCUMENTS

1. LISA Science Requirements Document, LISA Science Study Team, May 2018, ESA-L3-EST-SCI-RS-001 https://www.cosmos.esa.int/documents/678316/1700384/SciRD.pdf
2. LISA Science Operations Assumptions Document, U. Lammers, A. Petiteau, D. Texier, P. McNamara, O. Jennrich, ESA-LISA-ESAC-SOC-AD-001, iss. D, rev. 2.2, 20 May 2019.
3. LISA: A proposal in response to the ESA call for L3 mission concepts, LISA Consortium by Lead Proposer Prof. Dr. Karsten Danzmann, January 2017, https://www.elisascience.org/files/publications/LISA L3 20170120.pdf
4. Maximizing U.S. Participation in LISA Science: preparing for the opportunities and challenges of a novel flagship mission, NASA LISA Study Team, report submitted to the Astrophysics Division of NASA Headquarters, February 2018.

## 2. SCIENCE COMMUNITIES AND INVESTIGATIONS

This section responds to the first task in the Charge to the Augmented NLST (cf. Appendix A):

> **Identify the US communities most likely to use LISA data for scientific investigations.**
> - What kind of research projects and related activities will they be likely to do?
> - What are their anticipated needs?
> - Is there any precursor science that should be supported ahead of the launch?

There are many ways the US communities are likely to use LISA data for scientific investigations. Below, we enumerate users by scientific expertise. In our survey we also examined communities defined by their familiarity with LISA or GW data (e.g., LIGO, pulsar timing arrays); see Appendix B. Note that we have taken an expansive definition of 'using LISA data for scientific investigations' that includes use by education professionals and citizen scientists. This choice is motivated by the experience of LIGO/Virgo through GCN alerts, citizen science projects, and the results of our survey.

The following US communities are most likely to carry out scientific investigations with LISA data and the concomitant science products:

- **Theoretical Astrophysicists:** Scientists who use LISA data to constrain, guide, and develop theoretical models for astrophysical populations, GW signals, and the physics and phenomenology of individual source classes. They will need data products at various levels of fidelity that can be compared to simulations or theoretical models, or used to validate predictions about observations. In turn, they will propose additional physics or improved descriptions of sources to improve the data analysis.
- **Computational Astrophysicists:** Scientists who create high-fidelity computer simulations for use in astrophysics and gravitational wave analysis. Currently the gravitational wave community consumes numerical relativity results to develop gravitational waveform templates and search algorithms, resulting in a highly necessary and very active area of pre-launch LISA science (it has its own category below). Computational astrophysicists in this category use simulations to model, e.g., galaxy formation, dark matter, neutron-star mergers, and black hole dynamics on galaxy scales. Once LISA data are available, this community will need high-fidelity data to constrain poorly-understood physics within the simulations; this is very similar to the supernova simulation community working to model observed supernovae properties.
- **Gravitational-wave Astrophysicists:** Many gravitational-wave astrophysicists are working within the LISA Consortium, from the pulsar timing array collaborations, and the ground-based-detector collaborations. However, there is a nascent, but rapidly growing community of gravitational-wave scientists working outside the major collaborations. These groups pursue research using public data; they are functionally equivalent to researchers in traditional electromagnetic astronomy who work on archival data, and have similar needs for documentation and access. During the LISA flight era, they may well be a major LISA user group doing analyses that are similar or directly equivalent to those produced by the core science team, but with their own unique ideas and techniques. GW pipelines developed outside major collaborations can be valuable drivers of GW science, since they are instigated by a source independent of the mission. Ensuring data products have appropriate documentation will ensure this community can produce the best quality science with LISA data. Colleagues who observe in other gravitational-wave bands (both higher frequencies covered by ground-based interferometers, and lower frequencies covered by pulsar timing arrays) will use LISA data both to look for counterpart events (as with the stellar-origin black holes that begin their lives in the LISA band and merge in the LIGO band), but also to connect population properties for source classes that have sub-populations or evolutionary phases that radiate in different parts of the gravitational-wave spectrum.
- ***Observational Astronomers Focusing on Other Messengers:*** Astronomers using other messengers, including EM, neutrinos, and ultra-high-energy cosmic rays, will expand their science scope and technological capabilities over the next decade. These communities are expected to be dedicated consumers of LISA data, coupling their own observations to LISA data products in order to create broad, detailed descriptions of individual sources and transient high-energy events. Beyond mere consumers, this group will likely become key collaborators in constructing multi-messenger priors for LISA data analysis, following up alerts, and helping with joint studies of, e.g. sub-threshold events in archival data. Astronomers working on time-domain observations of tidal disruption events are a growing community and are, in general, interested in multi-wavelength and multi-messenger astronomy.
- ***Data Scientists:*** The uniqueness of the dataset, in particular the current concept of an evolving global Bayesian fit to the data, will be attractive to conduct research into the data analysis itself using, e.g. the latest innovations in machine learning. Data science is a rapidly-evolving field, but it is expected that discoveries in this field will require access to lower-level data products, perhaps even to the ATM level.
- ***Numerical Relativists:*** Numerical relativity waveforms are a critical input into the data analysis pipeline. Today the numerical relativity community is striving to produce high-fidelity waveforms to be used in LISA searches for SMBH binary mergers. In the LISA era, where SMBH binaries will be observed in the data with high and even enormous fidelity, the numerical relativity community may use these data to inform their own investigations about where their simulation efforts are inaccurate, missing specific physical effects, or need increased development efforts. The LIGO experience has shown that the data spur innovations in NR waveforms -- e.g. higher mass sources seem to require higher-order modes to be included in the NR simulations. LISA may find unexpected eccentricity, spin-precession, a richer than expected ringdown, or a deviation from general relativity that will require access to pre-processed data to rule out instrumental effects.
- ***Instrumentalists/Technologists:*** LISA will be the first gravitational wave observatory in space, but it will not be the last. With LISA Pathfinder, a tremendous amount of knowledge was gained by monitoring the behavior and performance of the instrument, collecting data on noise and sources of noise, and building models of the behavior of the instrument and spacecraft together. Similarly, lessons learned from LISA will inform future space-based gravitational-wave detectors. Since LISA will be the first mission of its kind, it is critical that detailed data about the spacecraft behavior and environment be part of the data products, to inform research on technology and design of future generations of gravitational-wave observatories.
- ***Non-LISA Domain Scientists:*** A key view to keep in mind is the following: any community that currently provides input to LISA or makes predictions about what LISA might observe or how LISA might operate is a candidate user community for LISA data. A good example is the Space Weather community. Today, information about the solar wind flux is used to estimate the influx of cosmic rays into a LISA spacecraft, which drives charging rates. After LISA flies, data from

- observed charging rates, radiation monitors, and magnetometers may be desirable to the solar wind and cosmic ray community. LISA Pathfinder showed this brilliantly by producing a study of micro-meteoroids, a topic that has been of intense interest to the spacecraft community for decades both theoretically and experimentally.
- *Citizen Scientists/Amateur Astronomers*: Mobile device apps and other citizen science opportunities reveal a surprisingly large engagement from engineers, medical professionals, industry leaders, and non-GW scientists. In our survey, approximately half are amateur astronomers. This group is interested in citizen-science initiatives, including EM follow-up opportunities, as well as educational material with a high-level of detail. These science enthusiasts constitute a substantial and sophisticated stakeholder community that is especially sensitive to policy choices.
- *Non-Scientists/Instructors/Public*: The general press, university public-relations offices, and teachers at both the university and K-12 level display a huge appetite for highly processed data products. For the general audience (like the press) the need is for data products that are useful in public settings, like audio-visual representations of events, sources, and graphics/diagrams/movies/tutorials about the LISA observatory itself. For the instructional community, data products in formats students can use (like text-formatted data files) and simple algorithms (e.g. python suites) that can be used in the classroom or instructional labs, are extremely valuable and are already in high demand. Public engagement and instruction help train prospective future scientists and inform the general public about the benefits and results of scientific exploration. Promoting public engagement and education should be an important goal for any publicly-funded scientific mission, and ultimately benefits everyone in the LISA user community.

## 2.1 RESEARCH PROJECTS AND RELATED ACTIVITIES

**What kind of research projects and related activities will they be likely to do?**

LISA will enable a wide variety of research projects for all of the user groups mentioned above. There is strong overlap in the research projects that different groups will pursue using LISA data. For example, LISA will be sensitive to MBHBs at very high redshift, which will inform all communities about the likely MBH seed masses. Theoretical astrophysicists will use this information to derive understanding of how MBHs form; computational astrophysicists will use this information as inputs and sub-grid physics for large-volume, high-resolution cosmological simulations. Electromagnetic observational astronomers will use this information to understand the high-redshift quasar luminosity function and identify specific objects for detailed case studies. We therefore organize the types of research projects in terms of the level of data product that are likely to be used to study a given topic. The range of scientific projects includes understanding the formation, evolution, and demographics of MBHs, advancing galaxy–black-hole coevolution, magneto-hydrodynamics, and accretion physics of multi-messenger sources, stellar-origin black hole population studies, white dwarf tidal disruption and white-dwarf binary mass transfer, measuring cosmic expansion from standard sirens, the nature of dark matter, exploration of gravitational physics of merging black holes, tests of general relativity including tests of the speed of gravitational waves, discovery of exotic phenomena such as cosmic strings, even space weather. In addition to the above, specific areas of investigation, there is broad discovery space for the unknown astrophysical sources and new physics.

The nature of LISA data demands a highly interconnected cycle of research. To find a source, a waveform must first be produced to use as a template in the global fit. That waveform requires a theoretical description of the source, a quantitative model of its behavior, and computation of the templates. If one of those templates produces a detection, that is, a fit with a sufficiently high SNR, then the cycle can recur repeatedly to test theoretical improvements, model improvements, and/or more accurate waveforms. Multiple researchers with different specialties could be involved in each iteration, starting *before* there is a first science product. Improved detection with one improved waveform will affect other detections, because of the presence of all sources in the data stream and the global fit.

The following list of example research projects and activities is organized by the levels of the data product on which those projects and activities will be carried out. The list progresses from those associated with catalogs, likely the most popular products, back upstream to the least processed products, which we expect that the fewest will use. This same organization will serve to respond to the question "What science is done with each type of data?" in §3.1.

### Catalogs (L3)

LISA catalogs will contain tens of thousands of individual sources, mostly Galactic binaries consisting of white dwarfs, neutron stars and stellar-mass black holes, but also stellar-origin black hole binaries seen to cosmological distances, massive black hole binaries ranging in mass from a few hundred to ten million solar masses, with mass ratios ranging from a million (Extreme Mass Ratio Inspirals, EMRIs), to one hundred (Intermediate Mass Ratio Inspirals, IMRIs) to less than 10 (MBHBs), stochastic backgrounds (galactic binaries, cosmological background, cosmic strings), bursts and possibly exotic sources (discoveries, cosmic strings, boson stars, etc). For individual binaries, the entries include source class (if known), masses (individual masses for evolving sources, chirp mass for slowly varying frequency), luminosity distance, spins, sky location, orientation of the orbital plane, possibly eccentricity, phase, time of merger, and possibly the waveform. Higher multiples (triples, quartets) and other uncommon systems are possible. The catalogs will also come with auxiliary data, such as the strain residuals after subtracting all identified sources.

**Learn the nature of individual sources and source types**
- Compare LISA source parameters with expectations from theory, EM observations, and/or combined multi-messenger/ multi-band observations for individual sources and source types.

- Extract new or better determinations of masses, spins, orbital periods, eccentricity, orbital inclination, orbital evolution, luminosity distance, etc.

- Study physical processes at work: mass exchange, tidal distortion, eccentricity, ringdown, the presence of higher multiples and possibly planets, the impact of circumbinary material on the GW inspiral. Compare simulations of source waveforms with fitted waveforms.

**Study the demographics of a source class**
- Compile and study the number density, mass function, spatial distribution, orbital period distribution of a source class, e.g., mass exchange binaries, stellar origin black holes, MBHBs.
- Map distribution of compact binaries throughout the Milky Way.
- Map distribution of MBHs throughout cosmological history, searching for seeds and growth mechanisms.
- Compare population information to predictions by stellar evolution models, galaxy simulations and cosmological simulations.
- Correlations of a source class with EM catalogs (including time-domain information)

**Characterize stochastic backgrounds**
- Measure the structure of the Milky Way using the astrophysical stochastic foreground of galactic binaries.
- Search for and/or set upper limits on conjectured backgrounds, e.g., a cosmological GW background, or that due to stellar origin black holes, and compare to theory.

**Learn about unexpected sources**
- Characterize and explain bursts - transient signals not carrying the signature of an inspiral.
- Search for and/or set upper limits on exotic sources (e.g., cosmic strings), and compare to theory.
- Study statistical evidence for unexpected sources.

**Search the residual strain data**
- Search the residuals for unknown source types, using statistical criteria.
- Use experimental templates to look for additional sources.
- Search residuals for evidence or upper-limits of backgrounds.
- Investigate improved templates from new physical models by adding back detected sources and refitting with the templates.

## [Alerts (L2)](#)

Alerts are low-latency notices of pending mergers, new detections, and sources evolving out of the LISA bandwidth and into the ground-based GW detector band in order to facilitate multimessenger/multiband observations.

- Carry-out EM/particle/ground-based GW observations for counterparts.
- Use multimessenger and multiband results to improve LISA global fit and parameter estimation.
- Combine information from EM and GW observations to improve deductions from both, e.g., surrounding matter, accretion disks, etc.

## Fitting Waveforms (L2)

Sources are detected by fitting a very large number of templates to L1 strain data. The suite of possible templates will be based on a high level of prior knowledge about potential sources. The data analysis process depends on theoretical and computational work to generate waveform templates for fitting. The template fitting process then identifies sources (discrete and stochastic backgrounds) and estimates the values of their parameters, with uncertainty estimates. While multiple approaches will likely be employed, the most promising strategy will employ an adaptive transdimensional global fit of source templates, instrument effects and noise models. The L2 product includes the source parameters of successful templates, their posterior probability distribution functions, and the templates themselves. Only those templates meeting robustness criteria are carried forward to the L3 catalog as 'detected,' and their templates are subtracted from the original strain data to produce the residuals. Backgrounds are treated analogously. The L1 data is well-suited for studies of data analysis, new waveform models, and consistency checks:

- Theoretical and computational studies to develop better waveforms for expected sources. This could be done by adding the source/source type back to the residuals and re-fitting (keeping all other aspects of the global fit the same), or by re-doing the global fit with the new waveform.
- Inject multi-messenger and multi-band information into global fit and study the consequences.
- Data analysis research in algorithms, global fitting, unknown source searches, recursion.

## Intensive Reanalysis (L0-L2)

The most demanding science results, like testing general relativity or verifying an important discovery, will require reviewing the entire data analysis chain from L0 to L3, and reviewing platform and payload housekeeping data. Any processing along the pipeline, e.g., time re-sampling, corrections, glitches, TDI, etc., could be subject to review. Projects and investigations at this level will require a deep knowledge of the spacecraft, the instrument, and the processing pipeline.

- Characterize and improve TDI, including interactions with clock noise and associated sampling noise and frequency noise.
- Study and characterize disturbances (e.g., acceleration noise effects) for measurement science and future mission technologies
- Study and characterize interferometry errors (e.g., displacement noise effects) for measurement science and future mission technologies
- Develop improved techniques for data correction, gap and glitch handling, noise reduction, temporal resampling, and data quality evaluation.
- Analyze possible instrumental effects contaminating low SNR sources, backgrounds and extraordinary discoveries.
- Make novel measurements of space weather, including observing the spatial structure of solar energetic particle events on the scale of millions of kilometers (https://iopscience.iop.org/article/10.1088/1742-6596/363/1/012045) in near real time. This requires prompt, public access to LISA magnetometer and radiation monitor data.

## 2.2 ANTICIPATED NEEDS

**What are their anticipated needs?**

### 2.2.1 Universal needs for the LISA Community

LISA is unlike any prior mission. The novelty of LISA data, the potential for discovery, humanity's first view of the mHz gravitational wave sky -- these are all indicators that the needs of the US research community will be substantial and perhaps atypical. Discoveries during the mission may well reveal unanticipated needs that would require agility in deploying support.

We elaborate on the data product access required for various LISA User Categories in §3.2. Broadly, however, we find that although highly processed catalogs and low-latency alerts will meet the data needs of the largest number of people, a significant number of science investigations and user groups require access to less processed and even PF data products.

Aside from data access, our User Survey overwhelmingly revealed the need for user support, with users expecting well-documented data, user-friendly public data pipelines, live and involved user support, added-value data products such as cross-matched catalogs, as well as training for researchers at all levels who wish to use LISA data for discovery and education. We cover some aspects of user support in §4.1.

Meeting the data needs of the US user community will require resources to reduce, store, and distribute data products, including running independent data-reduction pipelines within the auspices of the Consortium and experimental data-reduction pipelines through a peer-reviewed process such as a Guest Investigator Program.

Building research capacity among the user categories is a unique challenge for LISA. There is no formal graduate or undergraduate-level training program in gravitational wave or multimessenger astronomy and roughly 10 nodes of gravitational wave expertise; of these, ~30% reside in departments with little to no astronomy. There is not enough expertise to generate LISA data products, much less fully exploit the data. We discuss some capacity-building priorities in §4.2.

### 2.2.2 Needs enumerated by LISA research categories

Below we give a suggestive list of researchers' needs to carry out the research activities enumerated in the previous section. They are organized here by the same broad data product levels, similarly progressing from highest level/greatest demand (L3) to lowest level/least demand (L0 to L2). At each level, the access, tools, and support that are needed for the activities in §2.1 are sketched out. At this point in the definition of the science ground segment, these lists of needs can only be suggestive, not complete or even correct in all aspects.

## Research with catalogs (L3)

There are generally three kinds of activity with catalogs: searching for sources, including discrete sources, foregrounds, backgrounds and unmodeled sources; searching other source catalogs for counterparts; and performing demographic analysis on collections of sources.  ESA intends to serve a catalog of sources from its public archive, and some Member States are expected to offer mirrors of the ESA archive for their research communities. NASA may want to follow suit, and may consider a more extensive array of support functions, as has been the hallmark of NASA science centers.

**Access** – Users will need access to the following kinds of information from an archive.  The standard version of most or all these products are likely to be mirrored from the ESA archive.

- Catalog metadata: source summaries, processing history, data quality metrics: duty cycle, non-stationarity, potential source confusion, etc.
- Discrete sources: parameters and probability density functions, significance metric (e.g., SNR), waveform, model used to generate the waveform
- Foregrounds, backgrounds and unmodeled sources, including parameters and probability density functions, significance or upper limits, spectral and spatial characteristics, models used for templates
- Instrument models: instrument response models, noise parameters and probability density functions, correction parameters and probability density functions, correction models, noise models, gap and glitch models
- Time series: residuals, noise record, vetoes, glitches, gaps, sampling window
- Mock data: simplified catalogs, training catalogs

**Tools** – Users will need the following tools to conduct research with the archive.  Tools like these are commonplace at NASA science centers.

- Search tools for LISA Catalog, e.g., search by source type, parameter values, significance, counterpart information
- Tools for analyzing and displaying foregrounds, backgrounds and unmodeled sources
- Database tools to compile demographics on source types or characteristics
- Search tools for other catalogs: matching data across source catalogs
- Demographics tools: mapping, graphing parameter distributions of queried populations, comparison of parameters with surveys (Gaia, Rubin Obs.), etc.
- Tools for working with residuals: fitting non-standard waveforms, adding back sources, adding back instrument performance
- Tools for working with noise and instrument response
- Software library of catalog and analysis tools.

**Support** – There is considerable discretionary latitude in the level of support that NASA could provide to US researchers. These support mechanisms will all return more science for more funding, especially for users who only interact with the mission at the level of catalogs.

- Documentation: High level descriptions of mission, instrument theory of operation, GW primer, Introduction to LISA data analysis, Primer for accessing the archive, Primer on high level tools,
- Research grants: Guest Investigator, postdoctoral and graduate fellowships
- Training: prelaunch, online, classes, workshops, courses, at meetings, at Science Center, mock data, real data
- Online help: document, videos, use-case examples, tutorials
- Phone support
- Science Center visitor facility (offices, meeting rooms, etc) and long-term research/visitor programs
- User support consultants for in-depth and longer-term advice on pipelines/research. The XSEDE model allocates this bespoke support via a competed grant program.

## Research with Alerts (L2)

The general scenario for alerts is that the LISA mission will send out notices to subscribers of GW events for which near-coincident observations have scientific promise. Alerts can be expected to have less information, and less reliable information, about discrete sources than a catalog. The latency from data acquisition to alert issuance must be kept low enough for other observing assets to effectively respond.

The primary function of issuing alerts will likely reside with ESA, and NASA need not duplicate that "push" function, though LIGO/Virgo currently use NASA GCN system for their alerts. However, there will be ancillary functions where NASA may choose to offer a duplicate capability, or enhanced support, for use by the US research community.

**Access** – Users of Alerts need access to several databases related to Alerts. In addition to alerts issued by ESA, operators of electromagnetic observatories and particle detectors will want to search past alerts and associated information. The LISA data processing system will want to collect counterpart observations for use in reprocessing data with parameters determined by other observations (e.g., more precise sky location). These counterpart observations constitute a database to be searched by Alert users, Catalog users and researchers exploring improved global fits.

- Alert archive: source type, some parameters, waveform (e.g. phase), merger time, confidence measure (e.g., false alarm rate).
- Entering EM and particle counterpart and ground-based GW observations/upper limits
- Archive of counterpart observations (e.g., GCN circulars), notices and messages
- Rapid updates during protected periods
- Automated access to alert information

**Tools** – Alert users will typically search a region of the sky for a counterpart that matches the properties of the LISA source. The LIGO/Virgo experience shows that those observers can beneficially interact to narrow down candidate counterparts by the observations of others. This was strikingly demonstrated by LIGO/Virgo's first observation of a binary neutron star merger. With the much larger number of LISA sources, we anticipate that EM and particle observers will greatly benefit from tools to search a database of counterpart observations.

- Search tools for source alerts (e.g., GCN Notices), circulars and messages
- Search tools for counterpart observations
- Search tools for other catalogs, e.g., survey instruments
- Software library of search and query tools for automated access

**Support** – Alert users will have their own needs for support In their use of LISA alerts.

- Documentation: high level descriptions of mission, instrument theory of operation, GW primer, introduction to LISA data analysis, primer for Alert system, primer on search and query tools
- Training: prelaunch, online training, training classes, workshops and courses at meetings and Science Center, mock and real alerts
- Online help
- Phone support

## Research with Fitting Waveforms (L2)

The data analysis step from L1 to L2 looks for GW sources, instrument response, and noise sources by simultaneously fitting models to the TDI strain variables. The models of, say, discrete sources, are evaluated to produce waveform templates with specific parameters. Roughly, the fitting process searches for the set of templates that minimizes the residuals. The most successful set of templates becomes the catalog when detection thresholds are applied.

This global fitting process needs models for all processes that contribute to the strain data, whether they be astrophysical, instrumental or sampling. Astrophysics and fundamental physics theorists will want to experiment with, or replace, the models with alternatives that include more or different physics, for example. Numerical astrophysicists and relativists will research ways to improve accuracy and speed of the template calculations. Data scientists will explore the fitting algorithms and statistics of this global fit. All of these research activities will need to modify the components of the global fitting process to test their work. Research is carried out by global fits using experimental models, templates, and fitting algorithms. There will likely be analogous research on the alert pipeline.

**Access** – A basic tenet of LISA research is that all ingredients used to produce the standard produces will be made available to users.  Hence, users will need access to the models, the templates, the intermediate fit results and the algorithms, auxiliary data and software that were used to produce the standard L2 output.

- Standard models for discrete sources, backgrounds, foregrounds, generic unmodeled sources, instrument noise, instrument corrections, instrument response
- Standard astrophysical models (e.g., Milky Way, binary systems, galaxy catalog)
- Algorithms for generating templates (e.g., Effective One-Body approximation)
- Auxiliary information (e.g., correction input like spacecraft orientation or test mass actuation)
- Fitting algorithms
- Intermediate products from global fit (for diagnostic purposes)

**Tools** – In keeping with the tenet above, users will have access to the tools used to generate the standard products, both current and previous versions, and experimental versions.

- Modeling
- Template generation
- Fitting
- Diagnostic tools to evaluate models, templates, fitting – useful for sub-threshold searches
- Software library: numerical relativity and approximation codes, GRMHD codes, model codes, template generation code
- Tools for working with residuals, adding/subtracting instrument response, etc.

**Support** – Users at this level will be fewer, more sophisticated, and likely need higher levels of support from experts because of the specialized knowledge necessary to unravel the most complicated part of the LISA data analysis.

- Documentation: alert and standard pipelines, models, template generation, fitting
- Training: prelaunch, online, classes, workshops and courses at meetings and Science Center
- Science center visitor program
- Long-term collaboration teams with Science Center experts

**Computational facilities**, to support research on this computationally intensive part of the LISA data analysis.  A natural implementation would be a Science Center with the computational capacity for the normal production pipeline and a research pipeline.

### Research with Deep Searches (L0 to L2)

Researchers studying discoveries, exotic sources, tests of General Relativity, and/or subtle instrumental effects will likely want to closely examine the earliest processing of the LISA data stream for any defects that could alter the interpretation of downstream data products. This could involve reviewing any step that modified the raw data from the spacecraft and any auxiliary data that could aid in the diagnosis of unexpected instrumental effects or errors or weaknesses in the analysis pipeline. These research activities will demand the deepest knowledge about the flight system and the data analysis process. These activities could precipitate a demand for on-orbit tests of instrument performance or alternate analyses. As the experience with LISA Pathfinder indicates, L0 data could be posted on a public archive, but with very little documentation or support, and then only after the end of the mission when the SOC is no longer occupied by operations. We discuss the implications of this in section 3.4.

**Access** – NASA could mirror TDI variables (L1) as they are released by the SOC. Access to earlier products may be critical for some research activities, but may be restricted to Consortium members, or instrument team members.

- Platform and payload housekeeping data
- Data quality metrics
- Commissioning data

**Tools** – Very few researchers will engage in these deep diagnostic dives, and they may have such specialized needs that general tools could be of limited value.

- Searching, extracting, and analyzing auxiliary data

**Support** – Some members of research groups working on these research projects will have to be experts in instrumentation and the LISA flight segment.

- Documentation on flight segment design, implementation and test
- Connection to and collaboration with subject matter experts, for example for Science Center experts

## 2.3 PRECURSOR SCIENCE

**Is there any precursor science that should be supported ahead of the launch?**

Since LISA's data benefits a vast array of science fields, a number of US scientific communities will likely be active before launch preparing for its use and effect. Influenced communities will work to anticipate how LISA discoveries will alter their topic's research directions, and how their community will maximize their scientific benefit from LISA data. New techniques, data from observing other messengers (e.g., EM, neutrinos, cosmic rays), and novel research strategies will likely build from the current state of the art. Directions these communities take prior to launch will be difficult to anticipate fully. Below we provide a few broad examples of what we know now will significantly impact the scientific return on LISA data. We address topics from each of the US communities expected to be inspired by LISA data.

- **Theoretical Astrophysics:** Since LISA will observe in a new astronomical band, there is much left to be understood from basic astrophysical theory to better prepare for LISA data. Some theoretical work is essential in order to fully understand General Relativity's predictions (or those of alternative theories) for binary coalescence signals at the levels of precision needed to support LISA's high-SNR observations. There is much basic theoretical work to do to understand how SMBHs form, evolve, merge, and transform their galaxy hosts. Theory behind the formation and evolution of compact binary stars is immature and would benefit greatly from continued support. Theory will also be crucial in understanding the connections between LISAs GW observations and related observations in EM and other messengers. Areas where more theoretical work is crucial range include understanding the physical processes which produce coincident signals in GW and EM as well as how GW and EM populations are related.
- **Computational Astrophysics:** An important theoretical uncertainty is the appearance of SMBBHs in other messengers. To know what EM signature to look for, work on generating realistic simulations of luminous SMBBHs and their electromagnetic emission needs support. Computational/theoretical models and simulations of accreting binary systems are needed, which will require support for new techniques and computational infrastructure. Current expertise in the field reside in less than 10 research groups; without support, this field risks stagnating over the next decade. Fully coupling radiation and MHD together is in a nascent form, yet is critical for producing realistic EM signatures from pre- to post- merger. Simulation effort of compact binary stars and their genesis will enhance the impact the demographic data derived from LISA data will have on our understanding of stellar evolution. The spatial and mass resolution in cosmological simulations is much too poor to resolve the SMBHs and IMBHs expected in LISA. Support is needed to push the state-of-the-art simulations of SMBH/IMBHs and their host galaxies to better understand event rates and demographics.
- **General Relativity:** Modeling binary black hole inspiral and merger when the black hole masses are very disparate is beyond the reach of both numerical and general relativistic frameworks, but are essential to understanding Extreme Mass Ratio and Intermediate Mass Ratio Inspirals. Support is needed to push the limits of the systems able to be modeled accurately.
- **Gravitational-wave Astrophysics**: Since LISA will open a new band of astronomy there is a large amount of preparation needed to prepare the methods and tools needed to conduct these observations. The incoherent sum of several signals alters the way GW analysis has been performed so far. Discovering means for mitigating difficulties and performing accurate parameter estimation would significantly and directly benefit a large majority of LISA science.
- **Multi-band Gravitational-wave Astronomy:** Performing template searches of stellar-origin sources in the high-frequency end of LISA's band (i.e. trans- GW band sources) seems now to require an impossibly large number of templates. New GW search algorithms are needed to detect sources transitioning through GW bands (from mHz to O(10-100) Hz) quickly enough for alerting terrestrial GW observatories. Historical analysis of LISA data could be performed once terrestrial GW detections have been made, though this would reduce the scientific return of these sources.
- **Observational Astronomy Using Other Messengers:** LISA science will likely have its greatest value in conjunction with other astronomical observations. Astronomical surveys in advance of LISA will prepare the landscape for these observations, including EM observations of black holes across LISA's mass band, surveys of solitary and interacting galaxies to high-redshift, surveys of galactic binaries, and more. Some of this data will provide input for LISA analysis, allowing joint

EM+GW inference of parameters for some systems and providing auxiliary information to optimize the community response to LISA alerts.
1. The community of optical observers has been working for a long time to identify binary white dwarfs that can serve as verification sources. That work is extremely important and LISA science would benefit from its continuation. More binaries of a wider variety (e.g., white dwarfs plus neutron stars) with well-determined orbital parameters would be extremely useful.
2. A key precursor science task would be the search and identification of nearby SMBBH candidates.  Such candidates may be used to calibrate and verify LISA during its mission. Learning more about SMBBHs and their environments, in particular how they play in the galaxy development, would increase the impact of LISA detections once made.  In order to perform the search, multi-wavelength EM astronomers should be encouraged and supported to perform the searches, particularly because no strong candidates yet exist.

- ***Data Science:***  Searches with other messengers for LISA counterparts (e.g., compact binaries, SMBBHs) will entail sifting through archival data, high-cadence all-sky data (e.g., LSST, DES), and performing new observational campaigns, possibly following up on contemporary theoretical developments. Multimessenger astronomy is a growing field that will likely require substantial data science support -- not only to successfully discover these rare signals but to make inferences from heterogeneous data, complete with different cadences, biases, and systematics. One clear example: performing template searches of stellar-origin sources in the high-frequency end of LISA's band (i.e. multi-GW band sources) is thought to require an impossibly large number of templates.  LISA science would likely benefit from the efforts of data scientists to enable these searches.
- ***Numerical Relativity:***  The availability and quality of waveforms drive our ability to do LISA science.  Long before the launch data of LISA, waveforms provided by the numerical relativity community will be needed to prepare to detect and interpret the first SMBHBs.  Key questions that will be answered in the next few years include: *What accuracy do we need NR waveforms to achieve for LISA sources?  What are the likely astrophysical parameters and does the state-of-the-art NR waveforms cover these parameters to our desired accuracy requirements?*  Over the next few years, the NR community will be responding to LISA's needs with more accurate and efficient codes to answer the above questions.
- ***Instrumentation/Technology:*** As a new type of science instrument, observations will take place as the mission learns how to best operate and apply the instrument and its data.  To be best prepared for this work, scientists will have to prepare in advance with tools to explore and understand any instrumental effects which may impact the science data.  One example: generating the L1 data will require modeling and subtracting proof mass motion from e.g., thermal, laser, and internal gravitational noise, but these models are still in need of development.  Research in this area is especially crucial to support the potential for LISA to make grand discoveries which challenge scientific expectations, as the most important discoveries will make the strongest demand on understanding the instrument.  Finally, supporting this community beyond LISA will require continued research investments in GW technology.
- ***Non-LISA Domain Science:*** Further efforts exploring ways LISA data may benefit other science fields, and would be a broad impact to the greater scientific community.  Supporting this endeavor could also elicit dialogue with new scientists who may offer novel insights to how LISA data is used, which would enhance LISA's scientific impact.  One example from the LISA

Pathfinder is space weather; pulsar timing GW experiments yielded a new solar system ephemeris; an emerging example from ground-based GW detectors is seismic metamaterials.
- ***Citizen Science:*** Engagement with the community will most likely benefit LISA science returns and provide broad impact to society.  One could imagine that LISA data analysis pipelines could be offloaded to machines hosted by citizens, similar to Einstein@Home.  Readily accessible and effective educational materials on LISA, LISA sources, and how LISA data are analyzed would minimize the threshold for involvement from this US community. Development of these systems and/or other ways of involving citizen scientists would benefit US LISA communities. Precursor effort could involve brainstorming novel ways to involve US Citizen Scientists and developing materials and toolkits for this community.
- ***Education and Public Outreach***: Building a community around LISA science can start well before launch with a broad and systematic effort to educate the next generation of astronomers. This effort could include custom curricula in undergraduate and graduate programs that aim to prepare the next generation of researchers by giving them the appropriate combination of formal science background and research skills. In parallel, outreach efforts that reach the general public (adults and school students) have worked well for other missions, such as Hubble. For LISA, such an effort could inform the public of the exciting potential of multi-messenger astrophysics in general and the combination of gravitational wave observations and other messengers in particular.

# 3. THE VALUE AND UTILITY OF LISA DATA PRODUCTS

This section responds to the second task in the charge (cf. Appendix A):

> **Identify the likely LISA data products. Assess their scientific value and utility to US communities.**
>
> - What science is done with each type of data?
> - What kind of data are needed by the various categories of LISA users?
> - How is scientific opportunity affected by access to data at different levels?
> - What are the impacts of latency?
> - What are the pros and cons of proprietary time for LISA data and LISA data products?

For this second task, we use the LISA data product definitions as summarized in §1.3.

## 3.1 SCIENCE WITH EACH TYPE OF DATA

**What science is done with each type of data?**

The kinds of research projects and related activities enumerated in the answer to the first question of the first task (see §2.1) essentially constitute the type of science done with each major type of data. As such, the answer to that question is organized to be the answer to this question. However, a few additional comments are merited.

The natural starting point for the majority of users will be catalog data, the most highly processed science product. Our LISA Interest Survey shows that researchers are initially most interested in source catalogs. This is the data product most accessible to researchers that are unfamiliar with LISA or GW detectors in general. However, we anticipate that more sophisticated users will want to use data products from earlier in the analysis pipeline, and experiment with the pipeline steps that produce them.

## 3.2 DATA NEEDED BY USER CATEGORY

**What kind of data are needed by the various categories of LISA users?**

In short, any community that produces information for LISA is a candidate user community for LISA data in the future. It is plausible that the precision and fidelity of LISA data will produce a renaissance of development activity in many fields that today are not considered "gravitational-wave astrophysics."

Beyond the field of specialization we focus on here, an important lens to view data products through is related to prior professional training. The needs for the quality and fidelity of data are different for full-time researchers, tenured professors, postdocs, graduate students, undergraduates, and non-domain scientists/amateurs. All might need LISA data in some form to conduct research that might be geared toward writing papers or proposals, instruction, or building skills and capabilities that will be useful for becoming a participating member of the core LISA community. In this instance having **multiple access points to data**, and **data in various levels of processing and sophistication** are important to allow all categories of LISA community to participate.

|  | PF | ATM | L0 | L1 | L2 | L3 | AUX | AV | Alerts |
|---|---|---|---|---|---|---|---|---|---|
| **Theoretical/Computational Astrophysicist** |  |  |  |  |  |  |  |  |  |
| **GW/GR Astrophysicist** | 1 |  | 3 |  |  |  |  |  |  |
| **Cosmologist** |  |  |  |  |  |  |  |  |  |
| **EM/Particle Observational Astronomer** |  |  |  |  |  |  |  |  |  |
| **Numerical Relativist** |  |  |  | 4 |  |  |  |  |  |
| **Data Scientist** |  |  |  |  |  |  |  |  |  |
| **Instrumentation/Experimentalist** |  |  |  |  |  |  |  |  |  |
| **Non-LISA Scientists/Citizen Scientists** | 5 |  |  |  |  |  | 5 |  |  |
| **Educator/Public/Amateur Astronomers** | 2 |  |  |  |  |  |  |  |  |

Table 1: Data Levels various LISA communities may need to conduct the activities we discuss in the report. Anticipated fraction of each user category is denoted by hue, with darker hues indicating a larger fraction of users in each category. These are based on use-cases developed by the NLST.

Notes:

1. Scientists such as GW/GR Astrophysicists seeking to test fundamental theories may need access to probe lower-level data (PF/L0) working in conjunction with expert instrumentalists to jointly understand whether interpretations such as "new physics" or "unanticipated instrumental behavior" better explain observations.
2. As a public science project which cannot be easily reproduced it will ultimately be crucial that all scientific results are accessible to enable external scientific review. This may require that data at all levels, which has been found relevant in some aspect of science interpretation should be available for external reproduction of the science analysis together along with any documentation and essential tools.
3. Some multiband GW science may depend on review of L0 data, perhaps including reassessment of inferred instrumental parameters (eg TTL/ranging) jointly with constraints from associated ground-based GW observations.

4. Numerical Relativity will need access to L1 data to work before a full global fit has been computed to determine if the physics in the NR simulations (or other models of SMBHBs) are correctly represented. Tests of the validity of general relativity and physics missing from the simulations or models will require the isolation of the suspected SMBHB event in the L1 data stream such that all residuals between the data and the theory could be interpreted as missing physics.
5. Ancillary science, such as mapping the solar wind and magnetic field, will likely need access to direct unprocessed instrument channels, pre-flight information about the sensors, and knowledge of the data reduction pipeline and residuals

## 3.3 SCIENTIFIC OPPORTUNITY AND DATA ACCESS

**How is scientific opportunity affected by access to data at different levels?**

Below, we categorize the importance of each level of data to fully accomplish each science objective, with the darker hue being more critical.

|  | ATM | L0 | L1 | L2 | L3 | Alerts | AV | AUX |
|---|---|---|---|---|---|---|---|---|
| **SO1** GALACTIC BINARIES |  | ░ | ░ | ░ | ■ | ■ | ■ | ■ |
| **SO2** MASSIVE BHs |  |  | ░ | ░ | ■ | ■ | ■ | ■ |
| **SO3** EMRIs |  | ░ | ░ | ░ | ■ | ■ | ■ | ■ |
| **SO4** STELLAR ORIGIN BHs |  | ░ | ░ | ░ | ■ | ■ | ■ | ■ |
| **SO5** GRAVITY | ■ | ■ | ■ | ■ | ░ | ░ | ░ | ░ |
| **SO6** COSMOLOGY | ░ | ░ | ░ | ░ | ■ | ■ | ■ | ■ |
| **SO7** GW BACKGROUND | ░ | ░ | ░ | ■ | ░ | ░ | ░ | ░ |
| **SO8** DISCOVERY | ■ | ■ | ■ | ■ | ■ | ■ | ■ | ■ |

**Table 2:** Data Levels various Science Objectives may need to conduct the activities we discuss in the report. Anticipated fraction of each user category is denoted by hue, with darker hues indicating a larger fraction of users in each category. These are based on use-cases developed by the NLST.

**Catalogs (L3):** The most basic data product is a catalog of resolved sources, broken down by type of source, sky localization, and derived properties. For example, the entry for a resolved compact object binary would include sky coordinates, the modeled waveform, and information derived from the waveform (masses, orbital configurations, distances, etc). Binary black hole mergers, inspirals and other transient events would include event times as well. This information is vital to enable scientists to achieve several LISA Science Objectives.

A resolved source catalog would ideally be periodically updated to include new transient events and increased signal-to-noise data on known sources. For binaries with parameters that change little over the mission lifetime, updates approximately monthly would allow for appropriate calibration and prompt scientific response. For merging massive black holes, however, time scales are much shorter, and updates, especially of sky localization, should be given as often as feasible to enable EM localization and follow-up. Such rapid updates will likely result in a few missteps, including artifacts and calibration errors which will be corrected in future updates. The opposite extreme, a single catalog released at the end of the LISA tenure, would have the advantage of being clean and complete, but would fail to give the community the opportunity to follow-up on transient events in a multi-messenger or multi-band capacity, which could severely limit our ability to achieve science objectives 4-8. According to our LISA Interest Survey (Appendix B), a catalog of resolved sources is one of the highest priorities for the community. In fact, catalogs and high-level data are the most compelling reason for outsiders to join a large collaboration like the LISA Consortium.

**Added Value (AV) Catalogs:** increase in complexity and include cross-matches to EM counterparts (e.g., SDSS/LSST identifier). Such (AV) catalogs would facilitate scientific analysis of sustained sources such as compact object binaries, which will benefit from improved sky localization as LISA increases its operation time. For objects with more uncertain localization, such as massive black hole mergers, a catalog could provide a candidate list of EM counterparts, possibly weighted by probability based on galaxy properties such as redshift and stellar mass. While more labor-intensive, this increased information would greatly facilitate the identification and analysis of GW sources classified as compact/merging binaries. Respondents to the LISA Interest Survey noted that a catalog matching LISA sources with large EM surveys (e.g. LSST, Gaia) would be by far the most useful tool for the community's research. This indicates that AV material could engage the observational astronomy community by removing one of the biggest barriers to their follow-up of LISA sources.

AV catalogs in collaboration with EM/particle/ground-based astronomers can also aid in LISA data reduction itself. For those searching for GW counterparts to EM or astroparticle events such as, e.g. supernovae, flares, or gamma-ray bursts, a "multi-messenger priors catalog" would facilitate scientific analysis. In the event that LISA does not detect a GW counterpart, upper limits on event parameters would be useful to rule out specific physical models and/or environmental effects. This catalog would consist of input by the EM or astro/particle communities and would require a thorough search through L1/L2 strain data and/or detailed modeling of a predicted event, to determine upper limits on the observability of specified events at an input time, or to find sub-threshold events.

**Alerts:** Valuable to those focusing on time-domain studies including multi-messenger follow-up. The GraceDB[2] website run by the LIGO and Virgo collaborations is an example of such a catalog. For every LIGO/Virgo event (which includes sky localization information and generalized mass information), there is also a collection of multi-messenger alerts reporting on whether there was a cotemporal/cospatial event. Such collaborations have already borne scientific fruit (i.e., a binary neutron star merger with a corresponding gamma-ray burst) and would be extremely valuable for binary mergers detected with LISA. For such a collaboration to exist with maximal scientific return, LISA alerts would need to be issued as fast as is feasible. Associated EM events may be exactly co-temporal, so a delay could result in losing crucial multi-messenger data. Immediate alerts will run the risk of some retractions due to false positives; however, retractions need not cause undue strain on the community. In the LIGO example, alerts contain a false-positive likelihood, and retractions often happen quickly (within hours or a day). Such information and responsiveness are essential components of alerts. Retraction of alerts would be a necessary but small downside to the high scientific return of having immediate alerts in the first place. The LISA Interest Survey indicates that timely alerts are a very high priority for all types of users in the community.

**Beyond Catalogs:** *Some science objectives cannot be met with a catalog alone*. For example, SO 5 (Explore the fundamental nature of gravity & black holes), SO 7 (Understand stochastic GW backgrounds & their implications for the early Universe and TeV-scale particle physics), and SO 8 (Search for GW bursts and unforeseen sources) specifically require access to full measured strain data from the start of the mission to the time of analysis (which will likely be undertaken at several points during the mission) for a complete analysis. The data relevant to these investigations are not limited to specific events, but a combination of signals from numerous types of objects at all frequencies. Lower-level strain data are required.

**L0/ATM/PF data** could be used by scientists who are searching for low SNR measurements or unmodeled sources. A claim of new physics, for example, would require a deep dive into the calibration pipeline to account for background errors and to search for steps which may inadvertently filter out (or accidentally create) an unexpected astrophysical source. For example, if one optical bench creates a glitch through a relaxation mechanism near the time of an unexpected source, the resulting signals may be questionable and discarded from automated pipelines, even though the other 5 optical benches are working well allowing a re-analysis to create a valid candidate from the 5 benches. In addition, independent verification of results is enabled by the redundancy in LISA strain channels. Lastly, substantial incidental science will likely result from analysis of raw LISA data. For example, LISA pathfinder data has been used for a number of non-gravitational-wave science purposes, including to measure the galactic cosmic ray flux[3], monitor space weather[4], constrain the origin of zodiacal dust and micrometeoroids[5], and to constrain models of quantum mechanics[6]. Archival access and documentation for manual calibration would be required to make such data useful. According to the LISA Interest Survey, such information would mostly be utilized by those familiar with LISA.

---

[2] https://gracedb.ligo.org/

[3] https://arxiv.org/pdf/1711.07427.pdf, https://arxiv.org/abs/1802.09374, https://arxiv.org/abs/1904.04694, https://mdpi.com/2073-4433/10/12/749

[4] https://esa.int/Enabling_Support/Preparing_for_the_Future/Discovery_and_Preparation/ESA_s_unexpected_fleet_of_space_weather_monitors

--------------------------------------------------------------------------------------------------------------------

**To fully enable the community to accomplish all eight science objectives, at least L1 and AUX are required.** Data products at this level would likely be retroactively updated as the understanding of the detector improves. Providing L1 and AUX will allow for analysis of the stochastic GW background (SO 7) and searches for unmodeled events (SO 8). In addition, L1 and AUX will allow those testing the fundamental nature of gravity (SO 5) to re-analyze merger events and compare the fiducial waveform to those in modified gravity models. Strain data and residuals can also aid the search for deviations in predicted waveforms for any compact object binary/merger. For example, the existence of gas near a binary could speed up (or slow down!) a coalescence beyond what is expected by GW radiation alone, and an astrophysicist with a new interaction model could interactively re-fit the global fit solution to test and isolate its importance. L1 data are most useful to the community if there is sufficient documentation and interactive tools for its use, since strain is quite non-intuitive to scientists who are not already heavily involved in GW projects.

## 3.4 IMPACTS OF LATENCY

**What are the impacts of latency?**

In this section, we summarize the impacts that various levels of delay in event alerts would have on each Science Objective. Much more detailed information on each case can be found in Appendix D. For the purposes of this section, we define "latency" as the delay between the acquisition of data and access to a data product. This may occur due to data downlink from the spacecraft, time required to process data, spacecraft commissioning, data validation, or proprietary periods.

Latency periods will be different for users within the collaboration, for any potential data-sharing partners, and for the public. Currently, there are no estimates of processing times for L0, L1, L2, or L3 data, and, therefore, we consider generic latency periods, of any origin, in this section. It is important to note that latency periods are likely to be quite different for each level; for example, alerts would ideally have the lowest latency, while catalogs of sources and parameters would follow shortly after. Very low level data, such as L0, may require significant resources and incur a large latency period; however, access to these data is certainly important, especially for novel applications that require detailed instrumental and systematics mitigation.

Because the majority of Science Objectives are maximized by prompt electromagnetic follow-up, the science impacts of latency and different proprietary scenarios are deeply interconnected. For purposes of brevity, the impact of each objective is summarized here and in the accompanying Table 3. The details of different proprietary period scenarios are addressed in the next section; here, we treat proprietary periods as an additional possible source of latency. Note: the first two columns display latency expectations that may be unrealistic during standard operations in which data may be downlinked once per day.

---

[5] https://arxiv.org/abs/1510.06374, https://arxiv.org/abs/1905.02765

[6] https://arxiv.org/abs/1606.03637

|      | Few Hours | 2 Days | 7 Days | 1 Month | 4 Years |
|------|-----------|--------|--------|---------|---------|
| SO 1 | green | green | green | green | orange |
| SO 2 | light orange | orange | orange | red | red |
| SO 3 | light orange | orange | orange | red | red |
| SO 4 | green | green | light orange | orange | red |
| SO 5 | green | green | light orange | light orange | red |
| SO 6 | light orange | orange | orange | orange | red |
| SO 7 | green | green | green | green | green |
| SO 8 | orange | orange | red | red | red |

**Table 3:** Graphical summary of the impact of different latency periods on science objectives. Green indicates no significant impact on science return; light orange indicates a possible or unknown impact on science returns; bright orange indicates definite negative impact on at least one major initiative within the objective; red indicates severe impact/preclusion of at least one major initiative within the objective. We note here that satisfying the latencies the first two columns will likely require protected periods, but this was beyond the scope of our charge. We purely analyzed impacts on science.

### SO1: Galactic Binaries

Most galactic binaries detected by LISA will not experience significant change in their evolutionary state during the four-year primary mission. Therefore, the opportunity for electromagnetic observations to identify the counterpart will not be lost entirely, even with very long latency periods. However, simultaneous EM-GW observations, which would require source catalogs to exist while LISA is still observing, do have the potential to improve science return. For example: 1) Joint EM-GW observations drastically improve measurements of the distribution of binary orbital parameters, shedding light on evolutionary scenarios, and; 2) Repeated EM observations of GW-detected binaries over extended periods significantly improves the measurement of the orbital derivative, a critical parameter in studying tidal dissipation.

### SO2: Massive Black Holes

Merging SMBBHs are the fastest-evolving astrophysical sources among LISA's primary targets. The environment in which these mergers occur is critical to addressing this science objective, and electromagnetic counterparts are the key to such environments. This objective would suffer significantly from long latency periods and also from restricted access to sky-location alerts. Targeted EM searches must begin within hours of a MBH merger to have a good chance of detecting a counterpart. Additionally, the more observatories are participating, the more likely a counterpart will be found.

### SO3: Extreme Mass Ratio Inspirals

One of LISA's most unique and powerful capabilities is the ability to observe EMRIs, in which a less massive stellar origin compact object merges with a massive black hole. These events are excellent candidates for EM signatures due to the interaction of the lower-mass object with accretion flows around the massive black hole. The discovery of such a counterpart would have immense scientific value for a variety of applications, from star formation and cluster dynamics to gas cloud physics. The success of such a counterpart search, much like for MBH mergers, depends upon the ability to localize and follow up the event rapidly. Therefore, the science obtained from an EMRI detection would suffer significantly from high latency. Additionally, EMRI alerts would ideally include orbital phase information to aid in counterpart detection; verification of such information may add additional processing latency.

### SO4 and SO5: Stellar Origin Black Holes

Exploring the nature of gravity and understanding the astrophysics of stellar-origin black holes necessitates coordination with both EM observatories and ground-based GW detectors. LISA will be able to detect stellar-origin black hole binaries up to several months prior to their coalescence, meaning it can warn ground-based GW observatories like LIGO of upcoming events. Detections of such binaries by LISA and ground-based detectors enables black hole spectroscopy, a powerful potential test of deviations from general relativity. Any such deviations would be major discoveries.

The same ability allows LISA to provide EM facilities with sky localizations well in advance of merger. They will then be able to monitor the relevant region and obtain very prompt merger counterparts, providing valuable information on the close environment of these black holes. Thus, any verification latency that would prevent a pre-merger warning to ground-based GW detectors would be seriously detrimental to science returns regarding stellar-origin black holes.

### SO6 and SO7: Cosmology

For most science initiatives in this category, rapid EM follow-up of GW sources is not necessary. Delays on data release would have little effect on the science return regarding the stochastic GW background or for GW-only measurements of the dimensionless Hubble parameter. The only exception is the constraint of cosmological parameters through joint EM-GW observations; the considerations for this objective are the same as for the MBH mergers (SO2) above, because EM observatories provide independent measurements of the source distance.

Detection of a stochastic GW background will rely on extensive collaboration between theorists and instrumentation experts; therefore, unless a robust population of such theorists exists within the collaboration, it may be beneficial to limit latency for this reason in order to provide maximal independent verification of what would be a spectacular result. More details on how different proprietary models may affect claims of a stochastic background are provided in Appendix D.

### SO8: Unforeseen Sources

Perhaps without exception, the most groundbreaking discovery made by every new window into the universe developed by humanity has been unexpected at the time the technology was developed. It is impossible to predict what kind of EM, neutrino, or astroparticle counterpart to such a source might be expected, or what observatories would be best capable of performing follow up. Because of this, and the immense potential for groundbreaking discoveries from any unmodeled event detection, such events would ideally be publicly alerted as rapidly as feasible.

## 3.5 ANALYSIS OF PROPRIETARY TIME MODELS

**What are the pros and cons of proprietary time for LISA data and LISA data products?**

One reason to have a LISA proprietary period is to give the core team an opportunity to work on the data they labored to produce, as well as credit for the discoveries encoded within one of the most transformative observations of the Universe. Another highly practical reason for proprietary periods is cost. Historically, the most-used astronomical data releases (e.g., Hubble, SDSS) come with clear documentation and ample user support, and this requires significant financial resources and personnel; a short proprietary time would require more resources and personnel to release well-vetted, well-documented, and well-supported data.

We analyzed three broad data release models: Full, Partial, and Open proprietary periods. We note that in LISA there will be astrophysical sources detectable early in the instrument commissioning process and it may be that commissioning cannot be completed without running science analysis in parallel. For example, bright sources will need to be removed in order to get to a residual which can be compared with instrument models to assess the performance of the instrument. In the Full and Partial models here, this commissioning data is not released.

### [Full Proprietary Period:](#)

The most restrictive data access policy we consider in this document is a full proprietary period. Under such a policy, access to specific survey data products would be tightly restricted for some period of time. For example, alerts for newly identified sources could be restricted to collaboration members and selected observing facilities. Public release of newly detected sources prior to the end of the proprietary period could still occur in this scenario, likely in the form of papers or catalogs submitted for publication after internal collaboration review.

A principal advantage of a full proprietary period is that it ensures that scientists who commit to investing significant time in the project are fully recognized in the form of authorships on widely-cited peer-reviewed detection papers. Such recognition is most important for early-career scientists and graduate students. Therefore, restrictive data access policies can act as an incentive for scientists interested in working with LISA data to join the collaboration and commit to delivering specific analysis work products. Indeed, according to the LISA User Interest Survey, catalogs of resolved sources and access to alerts for follow-up observations were highly ranked as reasons to join a large collaboration.

A full proprietary period also allows a rigorous internal review process before sources are announced, potentially reducing wasted community effort as a result of false-positive detections. False positives are especially likely early in the mission, before unexpected characteristics of as-built detector data are fully understood and accounted for in data reduction pipelines. The commissioning period before the official start of observations can mitigate concerns regarding early-mission false positives, as the science and instrument teams can characterize the detectors without the need to issue alerts.

Even with a full proprietary period, user communities outside the LISA Consortium could still conduct population studies, compare catalog properties to theory, mine archival EM/Particle data to search for sub-threshold counterparts, etc., as long as the data and data products were well-documented and supported. Depending on the level of data product eventually released and its level of documentation and support, more significant science investigations could be undertaken, such as a re-reduction of the data with new tools or models.

One major disadvantage of a full embargo is that multimessenger follow-up can only occur after alerts are sent to observers. For sources with possible EM/particle counterparts occurring only hours or days after initial LISA detection, a complete embargo would substantially harm LISA multimessenger science opportunities by preventing prompt follow-up. The LISA Collaboration would sign data-sharing agreements with EM observatories, under which they could get access to alerts and share information within the collaboration, but would be bound by the terms of the embargo, and would leave out those user communities and observing facilities without a prior agreement. Such a strategy was adopted through 2017 by the LIGO collaboration, leading to the discovery of the electromagnetic counterpart to the binary neutron star merger GW170817a[7]. Note that LIGO has since transitioned to open public alerts due to the scientific and sociological disadvantages of the limited sharing approach.

Another major disadvantage is that since independent analyses aren't done, there is a missed opportunity for independent scientists to improve the data products during the mission by, e.g. discovering problems, folding-in priors from external data/events, or introducing novel analyses. This in itself could be another missed multimessenger/multiband opportunity in that potential target would only be discovered years after the event.

**Partial Proprietary Period:**

If the commissioning period is deemed insufficient to fully understand the range of detector issues, a hybrid option could be to enforce a full proprietary period, but only until a designated 'science-verification' period is complete to ensure inaccurate or false-positive detections are eliminated. Naturally, a partial proprietary period encompasses a range of data access and alert latency models.

---

[7] See the memorandum of understanding between LIGO and IceCube and the generic form agreement for EM observers for examples of such agreements

As one concrete example, the first year of data after the end of commissioning could have an additional science verification period, with public data releases embargoed until the period is complete. An ample science verification period would give time for both internal reviews of the data quality and publication of early high signal-to-noise sources, providing scientists who commit significant effort to mission preparation opportunities to write highly-cited detection papers. The Planck mission adopted this model.

Another concrete example could have a month-long science verification period combined with a low-latency public alert system and an expedited internal review process. As above, LISA will have built and tested its GW analysis pipeline during the verification period and produced the first published results. Therefore, collaboration scientists will have a distinct advantage over other groups hoping to analyze the data. At that point, if all past and new data are open access, external groups will be able to compete to extract, e.g., various signatures of a stochastic background. Such analyses may require the unique expertise of collaboration members. Therefore, the definitive, highly cited discovery papers will still ultimately be produced by the collaboration, regardless of any preliminary public alerts. Additionally, the public alert can be made citable, ensuring the collaboration is properly recognized.

Both models would reduce the rate of false-positives, though the short proprietary period would also reduce the time allocated to the Consortium to publish science results. We note that the degree to which any proprietary period incentivizes delivery of collaboration work products is strongly dependent upon the structure and quantity of funding available for the collaboration, including the data products and analysis tools provided to members. These aspects are discussed further in subsequent sections.

### No Public Proprietary Period:

In this model, data are released to the public immediately after verifying and validating that the instrument works. In recent years, the US-based scientific community has developed a general expectation that scientific data from publicly funded missions will be made public. Studies across a variety of fields have shown that open public access to the underlying data increases the number of citations and scientific impact of the research. Additionally, multi-billion dollar missions are generally selected for their unique capabilities. Consequently, independent replication of the underlying data by a second mission is often infeasible, given limited financial and workforce resources. Reproducibility is a core tenet of the scientific method, and essential for maintaining scientific integrity and public faith in the mission, agency, and the scientific community at large. Indeed, several large scale experiments have performed independent analyses, such as the Event Horizon Telescope, KiDS and DES reanalysis of weak lensing data[8], and LSST pipeline development applied to four cosmic shear surveys[9]. In communication with multiple PIs, there is substantial interest in full end-to-end independent replication.

---

[8] https://arxiv.org/abs/1804.10663.
[9] https://arxiv.org/abs/1808.07335

For a to-date unique and extremely sensitive mission like LISA, which is potentially vulnerable to a wide variety of known and unknown detector effects, full independent replication of results will require analysis pipelines that are able to work with every level of detector data, including the full spectrum of downlinked low-level telemetry. Engaging a wide range of experts in conducting and releasing these independent reanalyses as the mission is ongoing would be the best way to mitigate the science loss from transient events; this requires open and rapid data access.

Timing public release of data requires weighing both scientific and workforce concerns. It is in the collaboration's scientific interest to promote substantial community commitment to follow-up and precursor observations. Prompt, public alerts for newly detected sources maximize the number of observatories able to perform rapid EM follow-up. One concern with open public alerts is that they might reduce recognition for the collaboration and individual researchers involved. To mitigate this concern, alerts and data releases can be made citable and provided with persistent digital object identifiers (DOIs), ensuring recognition of collaboration workforce efforts in any external publications that occur before the official publication of sources and catalogs by the collaboration.

Many LISA sources are already visible with EM observations, namely the verification compact object binaries. LISA will benefit extensively from precursor observations by EM observatories, especially for characterizing the population of galactic binaries and cataloging potential host galaxies for sources. Several current and near-future observatories with substantial US-based contributions, including ZTF, LSST, and WFIRST, are already planning surveys to detect the EM counterparts to LISA sources. This use of LISA data represents a unique problem for data embargos. Data sharing agreements could be signed to reward observatories still collecting LISA-related data during LISA operations. However, some of the major scientific collaborations that will spend over a decade conducting observations that contribute substantially to LISA science will no longer exist by the time LISA begins collecting data. Therefore, the promise of easy public access to LISA data may better support communities that maintain substantial support for LISA science. Alternatively, LISA sources with counterparts already detected or otherwise present in pre-LISA archival data could be released immediately. Such a release would cite the observatory or observatories which discovered it, serving as a reward for communities that invest substantial effort in LISA precursor science.

If a source is first detected by LISA, the effect of public access on counterpart searches depends in part on how well-localized a source is. For well-localized events (~1 deg$^2$), or after a probable EM counterpart has already been identified, the number of observatories potentially capable of making observations of an event to good depth in multiple bands will be large. Consequently, for well-localized events, maintaining data sharing agreements with only a limited number of partners reduces the likelihood of a significant counterpart being discovered and thoroughly characterized compared to open public alerts. In fact, for such well-localized events, there is an appreciable probability that amateur astronomers could meaningfully contribute to counterpart searches, which is both of scientific value and an opportunity for public engagement, and is only possible if alerts are made public.

For poorly localized events (~100 deg$^2$) the number of observatories that individually have a reasonable probability of detecting brief faint counterparts is more limited. In such cases, relying exclusively on explicit data-sharing agreements with specific observatories may result in a relatively limited loss of discovery potential, largely because counterparts would have only a marginal chance of being discovered at all. On the other hand, having a larger number of observatories with different latitudes and weather conditions participating would increase the robustness of the search, and improve the fraction of the area that can be covered, especially if many observatories coordinate observations cooperatively using public tools, as is possible after a public alert. While the collaboration could consider the option of making alerts public only after the localization volume is sufficiently small, permitting external observatories to make their own decisions about whether or not to follow up a poorly localized event is likely to increase their overall engagement with LISA science, and therefore the probability a counterpart is found.

--------------

Regardless of whether or not there is an initial proprietary period, the scientific community expects all data and data products to eventually be made public and be well-documented to maintain scientific integrity. If only derived data products such as the computed strain are made public, community trust in LISA data will be reduced. A loss of trust in the mission may lead to both reduced scientific impact of published LISA results and reduced willingness to dedicate resources to cooperating with LISA in follow-up campaigns. Maintaining public archives also reduces the possibility of underlying data loss, ensuring LISA data are preserved for the future, when new analysis techniques and research questions are sure to be discovered.

Where independent replication of the underlying data is impossible, it becomes essential for scientific integrity that the data analysis process can be independently replicated, with different tools and analysis assumptions. One famous example is JADE (see e.g. https://arxiv.org/abs/1009.3763). Reanalysis of JADE data produced measurements of the running of the strong coupling constant an order of magnitude better than what was possible at the time, and better than any experiment since. Because the collaboration did not produce a public data archive, the fact that the data was not lost and the re-analysis was even possible was pure luck, and it did require extensive expert input. While it was extremely fortuitous that the required experts happened to still be available, it is certainly not something that any future multi-billion dollar taxpayer funded experiment should bank on in preserving its data for posterity.

If ESA ultimately chooses to have a full or partial proprietary period for some or all data products, clarity in data access policy is essential to maintaining community support for the mission, as it allows all stakeholders to make plans based on the availability of specific data products. To that end, all policy documents related to data access policies and proprietary periods, including a project data management plan, must be readily and publicly available on the NASA website, and include clear timelines for when specific data products will become publicly accessible. Data policy-making should carefully weigh the interests of all LISA stakeholders and science cases. Major changes to data access policies close to or during the mission are undesirable and can create workforce trauma.

From our analysis of the US user community, prompt public alerts are high priority data products, which requires a well-defined internal process for releasing a public alert rapidly, if such a release is deemed to be of substantial public or scientific interest. Such a procedure could consist of either empowering an individual with such discretion (e.g., the project scientist) or a small rapid response team with a defined set of representatives from different stakeholder groups.

|  | **Full** | **Partial, long** | **Partial, short** | **Open** |
|---|---|---|---|---|
|  | All data products released after mission | L3 released after year-long science verification period. Alerts released after 1 month. L0-L2+ AUX released after mission. | L3 released after few month long science verification period. Alerts released after 1 week. L0-L2+AUX are released in tandem with L3 releases. | All data products are released after commissioning |
| **Risks** |  |  |  |  |
| Consortium not able to do SOs | | | | |
| Consortium not credited for science | | | | |
| False alerts | | | | |
| Community not able to do SOs | | | | |
| **Benefits** |  |  |  |  |
| Cost-saving | | | | |
| Science Impact | | | | |
| Community Engagement | | | | |
| Scientific Integrity | | | | |

**Table 4:** Ranking of pros and cons for a range of data access models. Here, the darker hue indicates a stronger consequence in that category – for example, the largest cost-saving benefit occurs for a fully proprietary model, and the risk of the Consortium being unable to tackle all the Science Objectives is greatest if it must analyze and release all data after commissioning.

# 4. CONCEPTS TO SUPPORT US RESEARCH COMMUNITIES

This section responds to the third task in the charge (cf Appendix A):

> **Develop and assess concepts for how to most effectively support US research communities in their scientific exploitation of LISA data**:
>
> - What tools, documentation, and other resources should accompany the LISA data products identified above for each of these communities?
>
> - What interfaces should be made with other parts of the research community to enable further exploitation of the LISA data?
>
> - What model(s) of participation in the ESA-led LISA mission would best serve the US community?
>
> - Provide approximate timelines for initiating any activities discussed and prioritize them based on their importance for the successful scientific exploitation of LISA data by the US science community.

## 4.1 TOOLS, DOCUMENTATION, AND OTHER RESOURCES

**What tools, documentation, and other resources should accompany the LISA data products identified above for each of these communities?**

### ANALYSIS TOOLS

All LISA data products described above will need well-curated tools to maximize the discovery potential of those data. Data tools are designed to allow scientists to accomplish high-level tasks such as searching, classifying, reducing, and cross-matching data. The goal of well-curated analysis tools is to maximize the scientific output of LISA science for the greatest number of people, with a minimum of 're-inventing the wheel' at a cost to individuals and funding agencies. *Data analysis tools lower the entrance barrier for participation in and maximization of LISA science.*

**Tools to work with a LISA Catalog:** As described in the data products, the most basic LISA data product is a catalog of sources categorized by type, location, and derived properties. When designing tools for the catalog, it is important to take into consideration that the catalog is probabilistic: there is a probability attached to each event, similar to LIGO and Kepler. Providing basic tools to accompany such a catalog would allow users to search by category or source property (such as mass or sky localization), and plot the information. The search tools would ideally accommodate the improvement of source information with time. For example, it would be useful to include a feature for plotting the sky localization as a function of time or the evolving frequency of a merging supermassive black hole binary.

Astronomers would also benefit from tools that allow investigation of the residual data after loud, obvious sources have been fitted and removed. A simple catalog of detected events will not be sufficient for this task. For example, a user may wish to download the LISA data with the galactic foreground and MBHB signals removed, so that they can search for elusive EMRIs on their own. To do this, the data set would ideally be provided in its entirety, as well as with any resolved components and known foregrounds in a decomposed manner.

**Tools across catalogs:** According to the LISA Interest Survey, astronomers who do not consider themselves gravitational wave experts are very interested in matching data across source catalogs. The ability to mine source parameters from gravitational wave and electromagnetic wave data will facilitate the great discoveries of the next decade.

One can imagine a previously unknown source in EM astronomy being discovered through a correlation with the LISA catalog. While the flagship events (like LIGO/Virgo's binary neutron star GW170817) may be discovered without the aid of catalogs, more subtle inferences that require populations of events will be facilitated by the ease with which astronomers can move between catalogs.

**Tools for data reduction:** Although the ESA SOC will perform reduction of L0 data using software developed in-house, it is desirable that the same or similar tools be made available to the wider scientific community to facilitate maximum science return. Unfiltered, early-stage data products are crucial for those working on fundamental physics, such as tests of general relativity. Data that have not been filtered or cleaned are essential for novel science applications, since some filtering techniques to enhance detectability of expected signals may inadvertently wipe out signal from novel or unexpected sources. In order to use such data, full documentation of instrumental systematics and noise characterization are required, and software that extracts signals should be sufficiently well-documented for use outside the GW expert community by domain experts who may wish to incorporate new models or test new techniques.

It is likely that the wider community will eventually develop their own data reduction tools based on lessons learned from, for example, the LISA Mock Data Challenges, discussed further below. Support for these efforts is highly desirable for both maximal science return and the reproducibility of results. Other missions have benefited significantly from such user-developed tools; one prominent example is the Kepler mission, the impact of which expanded well beyond the detection of exoplanets due largely to the proliferation of open-access community-built pipelines[10].

**Tools for modeling:** Common questions among those new to the LISA community are: is my favorite astrophysical phenomenon a LISA source? How will theoretical and phenomenological astronomers work with space-based gravitational wave astronomy? Providing modeling software for various sources and scenarios, such as the ability to inject fake sources into the global signal, is a compelling way to engage a broader community of theorists and phenomenologists. Such software could be as complicated as numerical relativity open source code or as simple as tools for computing the signal-to-noise ratio. These types of data analysis tools can be provided independently of catalog decisions.

---

[10] See the list of user-developed software for Kepler and K2 here:
https://keplerscience.arc.nasa.gov/software.html

**Tools for Participating in Gravitational-Wave Astronomy**: The LISA Interest Survey indicates that the US community places a very high value on mock data. Mock data challenges are an excellent vehicle for novices and experts alike to work with simulations of LISA data, to see if one can write software to use for their own purposes. Historically, mock data have been powerful tools in other missions, such as TESS, in which a mock data release allowed user-developed pipelines to be developed and tested so that they could be deployed almost immediately after the first data release. Researchers may wish to determine whether their favorite gravitational wave model can be found in the instrument data, characterize the noise or instrumental factors they need to account for, or find out how sensitive LISA is to a novel source or alternative theory of gravity.

There is a community of people in gravitational-wave astronomy that creates and runs a mock data challenge for LISA. Current participants can engage by downloading the data and submitting entries using a web interface (https://lisa-ldc.lal.in2p3.fr/ldc). People who want to participate more vocally are welcome to join the telecons and mailing list – it is open to anyone, and does not require LISA Consortium membership. Adding dedicated US-based community support staff could ensure community engagement tools such as the challenges are easily accessible and comprehensible to interested community members.

These mock data challenges are also useful to the public. Properly maintained and supported, they could be deployed as activities in schools and as challenges for general public via citizen science, in a manner similar to SETI@home[11] and GalaxyZoo[12].

**Leveraging Ground-Based Gravitational Wave Astronomy:** Ground-based gravitational wave astronomy has developed a community of users, including those that use the publicly available data from the LIGO Open Science Community. LIGO data are very different from LISA data. How do we leverage the knowledge a segment of the US astronomy community has about LIGO data analysis for use with LISA data? LISA data always contains many resolvable sources; LIGO data currently has detectable signals roughly once per week. LISA noise is very different from LIGO noise, and what one researcher considers noise is likely to be someone else's source. The timing of the detector is also different, with MBHBs lasting months in-band while the detector orbits the Sun. A set of tools that provides researchers with tools to handle the nuances of LISA data will increase the participation of the community.

**Support of tools:** As a US community, we want to publicize and help shape this enterprise. Currently, the participants in the Mock Data Challenges are experts in gravitational-wave astrophysics, testing their own data analysis pipelines.  NASA may be able to help foster the development of stable, well-maintained tools to engage novices and experts alike. Python has a stable, open source code base with a tremendous suite of popular third-party packages easily entered by inexperienced developers. One respondent to the LISA Interest Survey mentioned that a curated data analysis toolkit of Python scripts in the form of a library or a collaboration with Astropy[13] would greatly enhance the usefulness of LISA data to the wider astronomy community. Indeed, there are several examples of NASA missions employing python for public user software, including the PyKe package for Kepler[14] and the Fermitools package in Conda[15].

---

[11] https://setiathome.berkeley.edu/

[12] https://www.zooniverse.org/projects/zookeeper/galaxy-zoo/

**DOCUMENTATION, TRAINING, USER SUPPORT:**

The most basic level of science support would consist of well-maintained documentation of data characteristics and data reduction procedures that describe any products released within or beyond the collaboration. Respondents to the LISA Interest Survey rated documentation as the highest priority for support resources. This is a basic resource that the entire astronomical community assumes will be provided for any mission data. Multiple levels of documentation should be provided for every resource that is made available to the scientific community, including a high-level overview, a detailed description of how the data product/analysis tool is to be used, and in-depth information about the detailed parameters associated with that product/tool. In addition, procedures for common analyses should be provided with sufficient supporting documentation that a new student can use the content to learn how to perform the most common analyses.

Respondents to the LISA Interest Survey also rated pre-recorded video tutorials on data reduction as a high priority resource. This can be considered an extension of documentation, and could be easily generated by filming in-person data analysis tutorial sessions.

Another important and widely expected support resource is a help desk, by which users can request assistance accessing data products, using data analysis tools, or understanding any aspect of the documentation via a dedicated email account. Ideally, the account would be monitored sufficiently well to enable replies within one day of a query; this is often done on a rotating basis by mission support staff. With higher levels of funding support, one can imagine a more interactive, rapid-response help desk format, such as live chat or Skype services. At an intermediate level between these two scenarios, it is possible for the mission to hold dedicated "office hours" during which live help can be obtained. This model has been used recently to acclimate users to the new format of the Astrophysics Data System[16].

An essential component of documentation and help desk support is the ability to provide user feedback which is responded to in a timely manner. Those using mission-developed public software should, for example, have access to an easy-to-use infrastructure for reporting bugs that provides real-time updates on the bug fix and patch documentation for new versions. Software versioning tools (e.g. gitHub) already provide such capabilities as part of their service.

---

[13] https://www.astropy.org/

[14] https://keplerscience.arc.nasa.gov/software.html#pyke

[15] https://github.com/fermi-lat/Fermitools-conda/

[16] The ADS office hours are now concluded, but can be found here: https://www.youtube.com/user/nasaads?fbclid=IwAR3G3ibgzehSGYOA0MORMauBVfGar5g2-B5sn0_e3Vhc3Yh_nqqdo9lfqpQ All were originally live-streamed and took user questions in the comments.

## OTHER RESOURCES

**Data training :** A major component of enabling science is training and informing new researchers in the procedures needed to analyze and interpret a new data set. Documentation development and training of interested scientists should begin well before launch through a variety of methods.

Devoting resources to data training in advance of mission launch will both increase the immediate yield of expected sources and diversify the science applications of LISA data. The LISA Interest Survey results indicate a strong preference for data training pre-launch, with 30% of respondents wanting data training as soon as possible, and 52% wanting it closer to, but before, the launch date. This was true regardless of familiarity with LISA data, indicating a broad enthusiasm that it would be wise to capitalize upon.

Training can take many forms. Online documentation and video tutorials can perform this function to some extent, but a more hands-on and effective method may be to support workshops at major conferences, stand-alone workshops, summer schools or hack-a-thon style meetings in which participants engage with simulated data under the supervision of GW signal experts. Hosting such workshops at existing astrophysics conferences would reduce the travel requirements of participants. However, it would also be desirable to host stand-alone workshops that provide travel support to participants from a wide variety of disciplines and academic institutions.

Earlier in this document, there was discussion of how LISA data may be used in public outreach, education, and citizen science efforts. Support for workshops to train educators on how to use LISA data for school projects, develop curricula points on GW science, and propose citizen science initiatives like GalaxyZoo would vastly increase participation in LISA data reduction and enhance mission popularity by engagement with the public.

**Mock Data:** Simulated data sets give researchers realistic data to analyze as they develop software tools and pipelines in advance of the availability of actual mission data. Mock data challenges, which have been discussed above in Curated Analysis Tools, can provide incentives to stimulate analysis methods and pipeline development. By gradually increasing the complexity of the data sets, the community can be informed on the variety of signals LISA will detect, the kinds of sources that can generate those signals, and the types of backgrounds that can affect the dataset.

Simulated data sets are an especially valuable resource for users new to GW data. Typically, simulated data will be generated in multiple stages with increasing levels of complexity. This means that, as users develop tools to utilize the data, they will be able to test those tools throughout the development process. However, the development of simulated data can be a complex process. It requires comprehensive knowledge of the signals that can be introduced into the system – instrumental, astrophysical, or otherwise. Such knowledge resides primarily in the instrument development team. For this reason, simulated data may not be generated and/or released to the community at large unless its production is specifically directed.

The growth of a new community of researchers can be enhanced by the sense of accomplishment that can be triggered by specific goals and a sense of friendly competition. Mock data challenges may provide incentive for development of tools and expertise leading up to the operational phase of LISA.

**Software and Algorithms:** Both mock data and real mission data will require software to analyze. In some cases, researchers will wish to develop their own software, in support of their area of study. However, software development can be time and resource intensive. A publicly-available set of standard analysis tools can open up the use of LISA data to a much wider community. Of course, such a toolset would require development, support, and maintenance to keep pace with technological development and resolve issues that are found. Of additional benefit would be the development and release of LISA-specific algorithms and data-handling modules as part of a commonly used analysis infrastructure, such as python 3. The existence of such code would drastically lower the barrier of entry for people developing their own analysis pipelines.

**Data portals:** A vital component of comprehensive scientific support for the broader LISA community would be portals with a single point-of-entry for accessing LISA-related data sets (real or simulated), high-level products, software, documentation, analysis guides/tutorials, and helpdesk. Satisfying the diverse needs of the community as described above requires a dedicated customized web presence that can respond to the needs of the users as they arise. One way to develop a broader community of capable users of LISA products is to provide a web-based interface to the analysis toolkit that would reduce the burden of software development.

## 4.2 Interfaces with the Research Community

**What interfaces should be made with other parts of the research community to enable further exploitation of the LISA data?**

As the US is a junior partner in this ESA-led mission, NASA's most important interface is with ESA. Likewise, as a member state in the LISA Consortium, the US has a responsibility to coordinate with the Consortium and support Consortium activities. To aid in data analysis, archiving and release, the expectation is that the US would host at least one of the Distributed Data Centers (DDC), and would coordinate data releases, including alerts, catalogs and pipelines; it is likely that the US DDC(s) would employ both the standard Consortium pipeline and, like many other DCCs, an independent one to aid in verification.

The US also has a responsibility to its other stakeholders. To maximize the impact of LISA data for Non-LISA domain science, NASA could interface with NSF, DOE, and NASA Divisions beyond Astrophysics to provide joint support for data reduction and novel science investigations. This may involve securing dedicated user support, computational resources, or hardware for these projects.

Likewise, facilitating multi-messenger work will require coordinating with major ground and space-based observational facilities to secure target-of-opportunity facility time during the mission. *Before* LISA launches, NASA could consider supporting EM observing campaigns of, e.g., verification binaries, to inform the global fit analysis pipeline. Beyond the major facilities, the interest and excitement of the amateur astronomy community could be harnessed if there were a coordinated effort to interface with the community, much like what has been done with exoplanet transit follow-up observations (e.g. KELT). Much of this support could be done with cross-agency coordination.

In the current gravitational-wave astrophysics ecosystem, driven largely by advances and discoveries from ground-based gravitational-wave observatories, there is strong demand for highly developed data products (catalog data, plottable extracted waveforms, public level summaries of scientific results, and audiolizations of gravitational-wave signals). These data products are being used by amateur astronomers, citizen scientists, the press, university public relations offices, and teachers. This diverse user community is a critical consumer of these high level data products because they are responsible for the broad dissemination of the scientific results and importance of gravitational-wave discoveries in modern astrophysics. Teachers are a particularly important target audience, as they are at the front end of the STEM pipeline that is producing the next-generation workforce that will be the scientists and engineers in the decades after LISA flies.

If the US wishes to maximize participation in LISA science, connections could be made with graduate and undergraduate institutions to support the development of the future LISA user community through deliberate early-career workforce development programs such as LISA Graduate and Postdoctoral Fellowships, visiting scholars and a LISA Colloquium tour. In addition, this community would benefit from user tools and mock catalogs to incorporate into undergraduate/graduate curricula, long-term workshops, etc. One successful model to broaden participation to underrepresented groups is to deliberately interface with, recruit talent from, partner with, and support Minority Serving Institutions to help establish research capacity at colleges and universities. (e.g., the Fisk-Vanderbilt Masters-to-PhD Bridge Program).

The US could take more significant steps to increase capacity in LISA science by implementing the recommendations from the 2019 NLST Science Support Taskforce report, e.g., faculty sabbaticals pairing traditional astronomers with LISA scientist partners, Faculty and Student Team Programs, cluster hires, and a ***flagship level LISA Science Center***, which could serve as a nexus to coordinate the activities we describe.

## 4.3 MODELS OF PARTICIPATION FOR THE US COMMUNITY

**What model(s) of participation in the ESA-led LISA mission would best serve the US community?**

The LISA data will be complex in structure and rich in astrophysical content, promising to return wide-ranging scientific results that will drive the frontiers of gravitational-wave astrophysics for decades to come. Maximizing the extraction of science from the LISA data will require large, collaborative teams with a diverse range of talents ranging from astrophysics, to data science, to gravitational physics, to computational engineering. For virtually the entire history of LISA to date, this diversity of talent has been drawn from both the US and the European scientific communities, integrated together in a large collaborative enterprise. In the LISA flight era, the continuation of this open, unrestricted collaboration will not only feel natural, but will also provide the best opportunity to preserve the expertise and institutional memory of people who have worked on LISA for decades or more.

There are models for US participation in ESA-led missions that restrict the numbers of US scientists who can be directly involved in the scientific collaboration around the mission. In the case of LISA, a restricted membership model would be detrimental for many reasons. First and foremost, a mission that has had the long development time of LISA will naturally have a large, engaged community of senior scientists and engineers who have spent a long fraction of their careers working on LISA. They will be natural representatives in a restricted scientific collaboration, but they would be competing for membership against more junior scientists who are attempting to build their careers (and the future of gravitational-wave astrophysics) on the LISA data. Such competition is detrimental to maximizing the science returns of LISA, and detrimental to fostering an open and collaborative ecosystem in science.

An ideal model for the broadest US participation would be an open access model, centered in a flagship level LISA Science Center. Such a center provides a focal point to preserve US institutional memory in LISA, could host and coordinate workshops and seminar series, and could provide a natural institutional structure to mirror and collaborate with our European colleagues who will be operating with a similar support structure in Europe. Establishing a US LISA Science Center *well before launch* can have a beneficial impact on the participation of the broader astronomical community by providing training, hosting topical workshops, disseminating mock catalogs, software pipelines, and documentation. Past experience indicates that successful science centers, like the Space Telescope Science Institute, are established several years before launch; this early adoption model may be especially relevant for a novel mission like LISA.

## 4.4 TIMELINES AND PRIORITIES

**Provide approximate timelines for initiating any activities discussed and prioritize them based on their importance for the successful scientific exploitation of LISA data by the US science community.**

### Top Five High Priority Activities: Near-term (2020-Mission Adoption)

**1. Begin formulation of the US LISA Science Ground Segment,** including a flagship-level Science Support Center. Develop the definition, design, and research/education/support activities. Curate a list of data products, detail its contents, and analyze infrastructure needs. This will need to be done in coordination with the ESA LISA Project and will likely involve a discussion of the data access, data flow, proprietary periods, latency issues, and support models touched on in this report.

**2. Support mission data analysis work packages**: we conservatively estimate at least 4 FTEs/year to support the development, documentation, and testing of the standard data analysis pipeline.

**3. Expand LISA Preparatory Science Program:** given the range and depth of precursor fundamental and data analysis research, this grant program could be accelerated to a yearly call and expanded in scope.

**4. Implement extensive capacity building and user training:** As mentioned in our 2019 Science Support Taskforce Report, the novelty of gravitational wave astronomy requires a deliberate effort to grow a new community of users. Measures such as Bridge Programs, funded multi-messenger research partnerships, faculty, postdoc, and graduate LISA fellowships, Faculty sabbaticals, student/faculty research teams partnering with LISA experts, workshops, faculty/postdoc cluster hires, user training seminars, formal curriculum, online tutorials, toolkits for new LISA users, etc, will be most effective if started early – this will allow the workforce development pipeline to grow in tandem with the data analysis and research pipelines.

**5. Develop prototype user tools:** The TDI simulator is an especially important tool for those building the data analysis pipeline, as is a waveform generator and prototype global fitter. For a broader user community, mock catalogs and visualization tools would help acclimate the future user base to LISA's capabilities.

# APPENDICES

**Charge to the augmented NLST**

October 7, 2019

**Introduction**

NASA is pursuing a role as a junior partner on the ESA-led LISA mission, which h**as an expected** launch date in the 2030s. NASA's contributions may include elements of the flight system, engineering support, and contributions to the science ground segment. In response to the NASA contributions, ESA will provide the US research community **access to LISA data and** opportunities to participate in LISA science. During the mission, NASA will support US scientists in their exploitation of the LISA data. NASA has begun early discussions with ESA regarding participation in the science ground segment and access to LISA data for the US community through the public release of data and data products.

NASA is charging the NLST, as proxies for the future LISA user community in the US, to provide scientific analysis to inform these discussions as well as the NASA policy and budget formulation processes surrounding LISA. The NLST is encouraged to collaborate with experts inside the NASA LISA Study Office and coordinate with European colleagues within the LISA Consortium, but should still retain their independent voice. This charge is broken into the following tasks.

**The Tasks**

NASA wants its discussions with ESA to be informed by factual knowledge about the **needs of** the US community for performing LISA science. To this end, the NLST is hereby tasked with the following:

**Identify the US communities that are most likely to use the LISA data for scientific investigations.**

- What kind of research projects and related activities will they be likely to do?
- What are their anticipated needs?
- Is there any precursor science that should be supported ahead of the launch?

**Identify the likely LISA data products and assess their scientific value and th**eir utility **to the US astrophysics communities**. LISA will produce a rich set of **data products at a** variety of levels (e.g., catalogs, individual system parameters with error bars, residuals from global fits, TDI "strain" data, lower-level instrument data, etc.). Issues the NLST will focus on include:

- What science is done with each type of data?
- What kind of data are needed by the various categories of LISA users?
- How is scientific opportunity affected by access to data at different levels?
- What are the impacts of latency?
- What are the pros and cons of proprietary time for LISA data and LISA data products?

While the information provided by the NLST may be used by NASA to inform data policy discussions with ESA, the NLST should focus on the scientific impacts.

**Develop and assess concepts for how to most effectively support the US research communities in their scientific exploitation of LISA data.** Building on the outcome of the previous task, the NLST will address the following:

- What tools, documentation, and other resources should accompany the LISA data products identified above for each of these communities?
- What interfaces should be made with other parts of the research community to enable further exploitation of the LISA data?
- What model(s) of participation in the ESA-led LISA mission would best serve the US community?
- Provide approximate timelines for initiating any activities discussed and prioritize them based on their importance for the successful scientific exploitation of LISA data by the US science community.

The NLST shall *not* provide a specific proposal for implementation of any activities (e.g., specific data right policy, buildings, computers, people, etc.) and will limit itself to the science impact of the findings.

To allow NASA enough flexibility during the discussions with ESA, the NLST shall consider various scenarios and their pros and cons. For example, what various capabilities should a LISA Science Center should have, with pros and cons of each? Findings will be used by the LISA Study Office and HQ for budget formulation.

**Deliverable due date**

The NLST will prepare a written report with the analysis and findings and deliver it to NASA by **February 1, 2020.** The report will remain sensitive and internal to NASA. A single PDF file will be submitted to the LISA Program Scientist, Dr. Rita Sambruna (rita.m.sambruna@nasa.gov), and to the LISA Study Scientist, Dr. Ira Thorpe (james.i.thorpe@nasa.gov), by the above date.

P. Hertz, APD Director                                                                                                      **Date**

# APPENDIX B - LISA INTEREST SURVEY

We report the results of the Laser Interferometer Space Antenna (LISA) Interest Survey. The survey was drafted by the NASA LISA Study Team (NLST) in response to a charge by NASA. The goal is to determine U.S. researchers' needs and interests in using future gravitational wave data from LISA. The target audience is the U.S. astronomical community. The survey contains 16 questions. Four questions were demographic in nature, ten questions asked for specific (rank-order) responses about gravitational wave data, and two questions were write-in.

The survey was posted on December 9, 2019, and ran through the annual American Astronomical Society meeting, ending January 10, 2020. Announcements were made on the American Astronomical Society blog and bi-weekly newsletter, the Astronomers Facebook page, the NASA Physics of the Cosmos listserve, and the Gravitational Wave International Committee listserve. Desiring a greater number of responses, the NLST decided to extend the survey to January 31, 2020, and re-posted the accouncements, additionally posting to the Sloan Digital Sky Survey listserve and the Large Synoptic Survey Telescope listserve. We received 193 responses as of February 1, 2020.

We had a good diversity of respondents from over 100 unique institutions, a broad range of career stages, and a good mix of theorists and observers. We present the results of each question below. The rank-order responses are plotted as sets of histograms: each response received a different number of 1, 2, .. , N ranks. The numbers do not always share the same sum, however, because null responses are ignored.

**1. How familiar are you with the gravitational wave mission LISA (the Laser Interferometer Space Antenna)?** (mark only one oval)

1.1 Very familiar.
1.2 Somewhat familiar.
1.3 Heard of it, but otherwise unfamiliar.
1.4 Never heard of it.

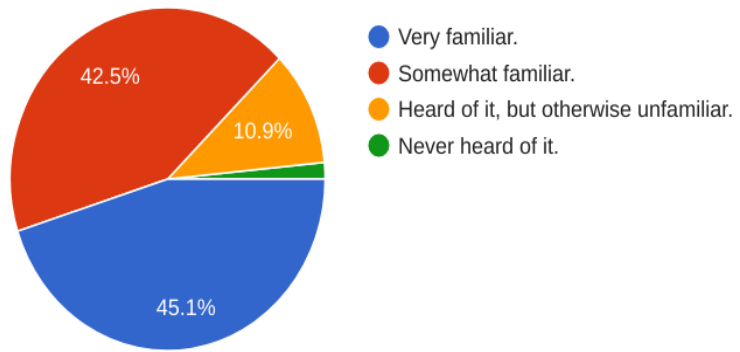

The first question presents an interesting result: approximately half of the respondents are "very familiar" with LISA, half are not. Since the goal was to survey the broader community, we split the remaining results approximately in half, by familiarity with LISA. Most results were common to both halves; the biggest difference was interest in low-level data (those unfamiliar with LISA expressed little interest in low-level data).

In brief, the highest-ranked data product was the catalog of resolved sources, followed by LISA alerts. For tools, the highest-ranked choice was a cross-matched LISA source catalog, followed by high-level analysis tools, and then recipes for common use cases.

Respondents overwhelmingly ranked written documentation as the top data-training choice, and ranked having a "general introduction" to LISA higher than other options. The majority of respondents thought LISA data-training should happen close to launch. A question about possible LISA-related initiatives showed no clear result; all choices received similar numbers of all possible ranks.

Interestingly, we asked about LISA data products twice (in questions 3 and 9). The second time we framed the question as "Would you be willing to join a large collaboration to use the following LISA data?" Responses are correlated with the first question about interest level, but half of respondents are unwilling to join a large collaboration to use LISA data.

We had two write-in questions, asking what precursor science should be done, and asking for final thoughts. We received dozens of responses. We print full set of results below.

**2. What is your familiarity with gravitational wave data?** (Choose all that apply!)

2.1 Familiar with LIGO/VIRGO alerts.
2.2 Familiar with LIGO/VIRGO data.
2.3 Familiar with Pulsar Timing data.
2.4 Familiar with future LISA data.
2.5 Not familiar with gravitational wave data.

### *Very Familiar* Respondents (N=87)

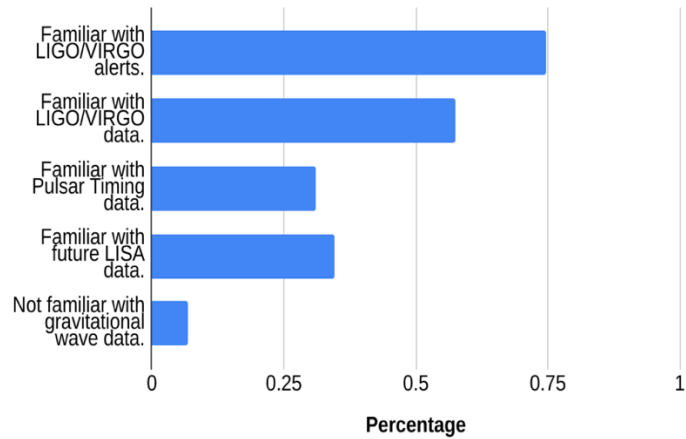

### *Somewhat Familiar* or *Unfamiliar* Respondents (N=106)

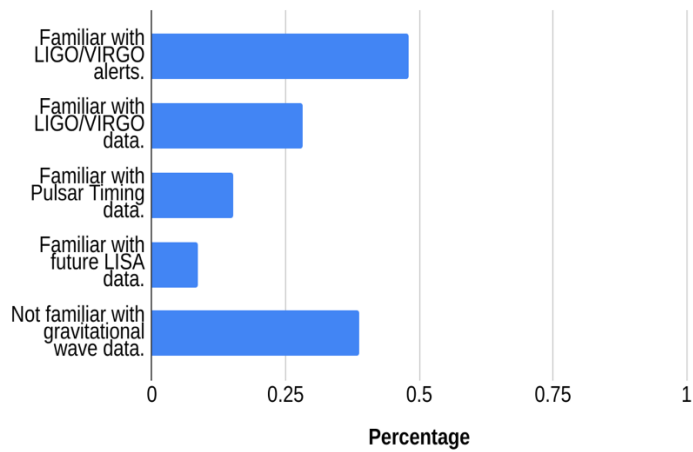

**3. What types of LISA gravitational wave data might be important to your research?**

(Please rank-order the options, where 1=most important and 6=least important.)

3.1 Alerts for follow-up observation.

3.2 Catalog of resolved sources, both continuous and transient.

3.3 High level data (e.g. all-sky strain maps) with Galactic foreground.

3.4 High level data with Galactic foreground removed, plus foreground model.

3.5 Frequently updated but less accurate 'quick look' data for monitoring potential sources.

3.6 Low level data with gaps, glitches, instrument noise, plus access to select engineering channels.

***Very Familiar* Respondents:**

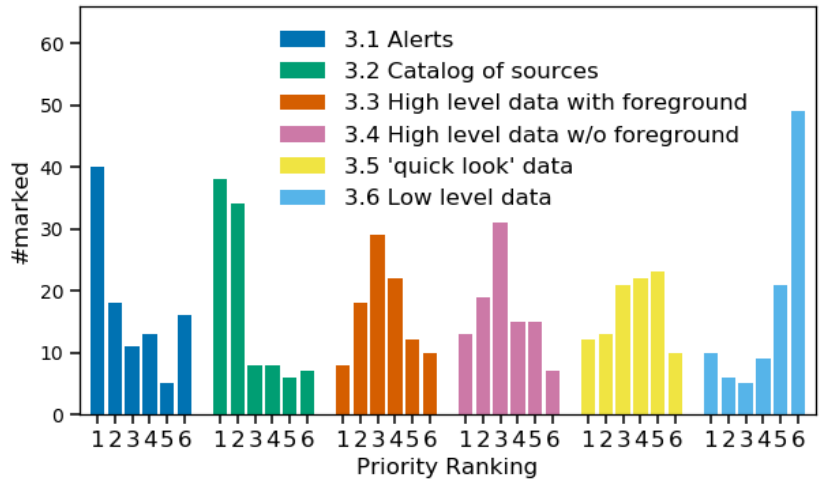

***Somewhat Familiar* or *Unfamiliar* Respondents:**

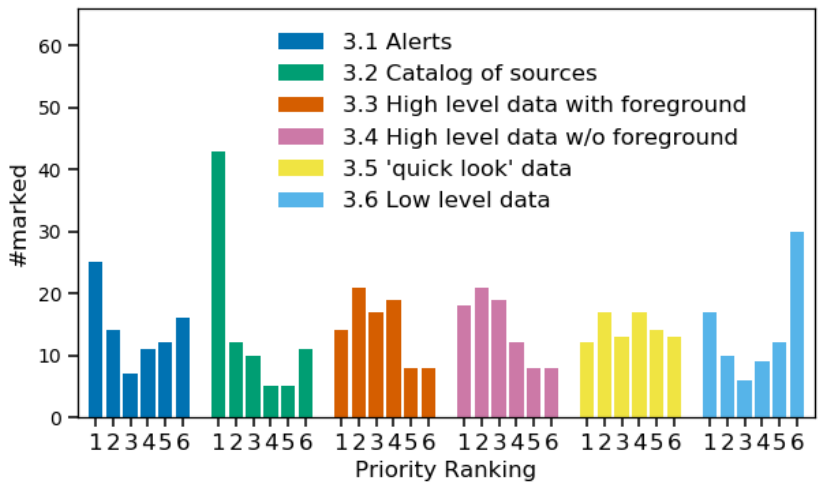

**4. If provided with tools to analyze mock LISA data, would you be interested in exploring mock data now?** (mark only one oval)

4.1 Yes.
4.2 Maybe.
4.3 No.

***Very Familiar* Respondents:**

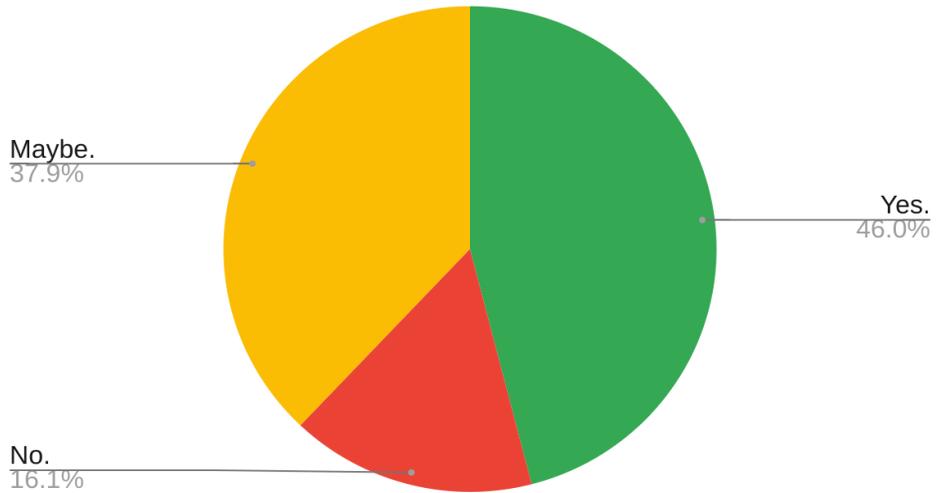

***Somewhat Familiar* or *Unfamiliar* Respondents:**

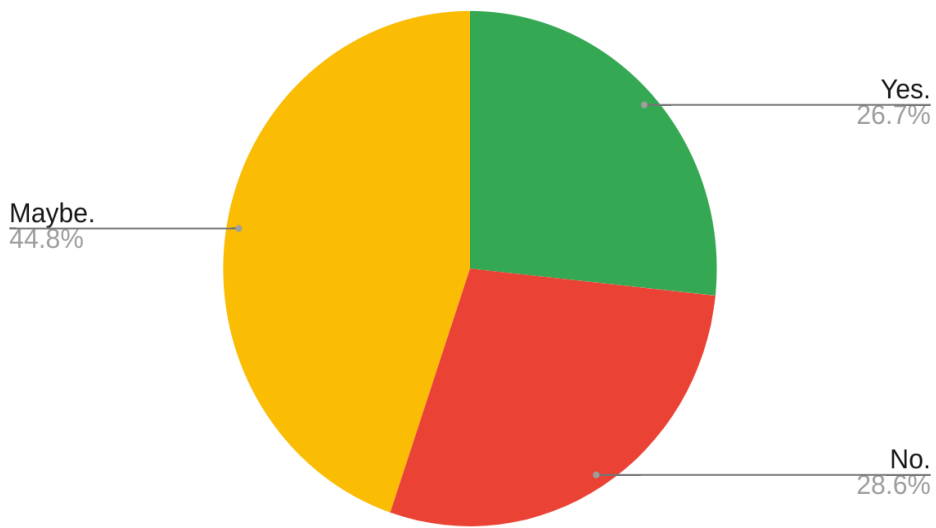

**5. How useful would the following tools be in enabling you to use LISA gravitational wave data, mock or real?** (Please rank-order the options, where 1=most useful and 4=least useful.)

5.1 Having the LISA source catalog matched with other major surveys (Gaia, LSST, etc)   .
5.2 High-level analysis tools, developed by the mission.
5.3 Recipes for common use cases.
5.4 Low-level data tools, with sufficient spacecraft information to determine if something is an artifact.

*Very Familiar* Respondents:

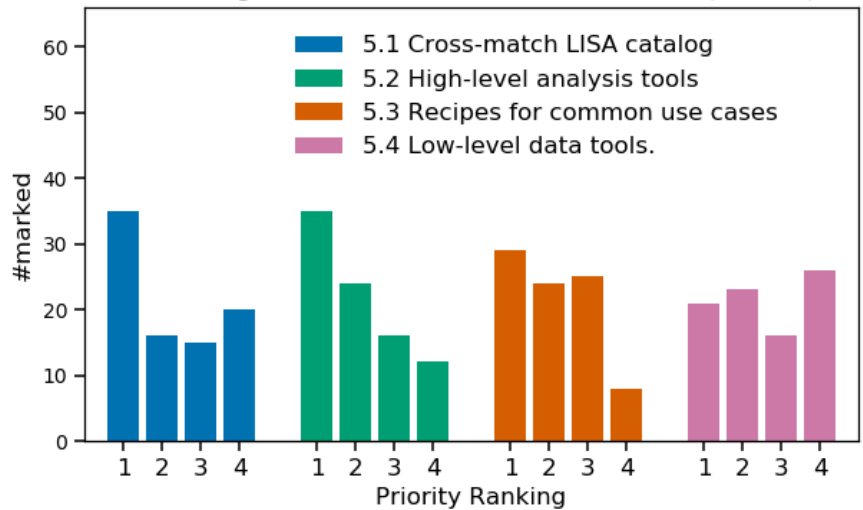

*Somewhat Familiar* or *Unfamiliar* Respondents:

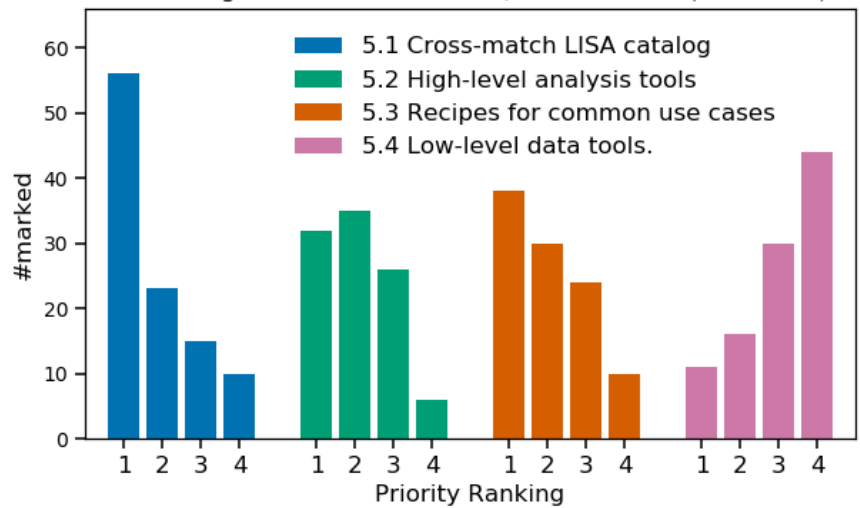

**6. To prepare for using LISA data, what types of LISA data analysis products are important to you?**
(Please rank-order the options, where 1=most important and 5=least important.)

6.1 General introduction to LISA and gravitational wave data.
6.2 Specific use-cases, with predicted errors.
6.3 Simulated source catalogs.
6.4 Simulated time-strain data.
6.5 Detailed mock LISA data analysis.

*Very Familiar* Respondents:

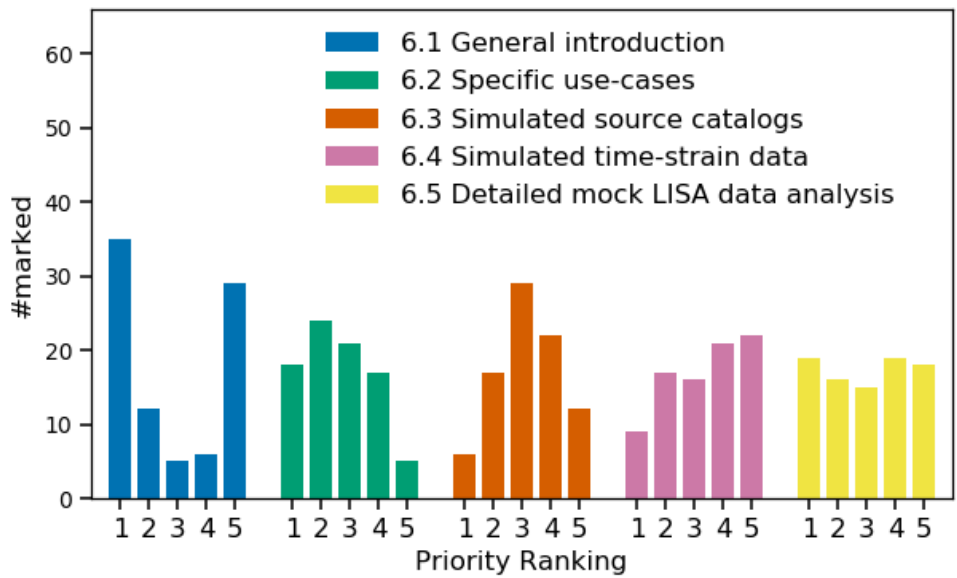

*Somewhat Familiar* or *Unfamiliar* Respondents:

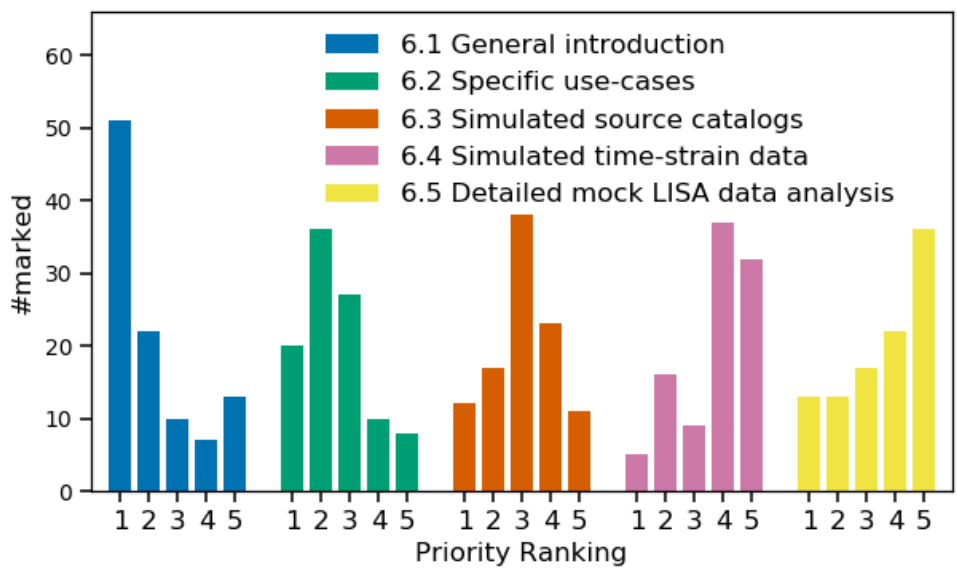

**7. How interested would you be in the following data-training options?**
(Please rank-order the options, where 1=most interested and 6=least interested.)

7.1 Written documentation and tutorials.
7.2 Pre-recorded video tutorials.
7.3 Online live training sessions.
7.4 Short workshops at existing conferences.
7.5 Summer schools devoted to LISA.
7.6 LISA mission help desk.

*Very Familiar* Respondents:

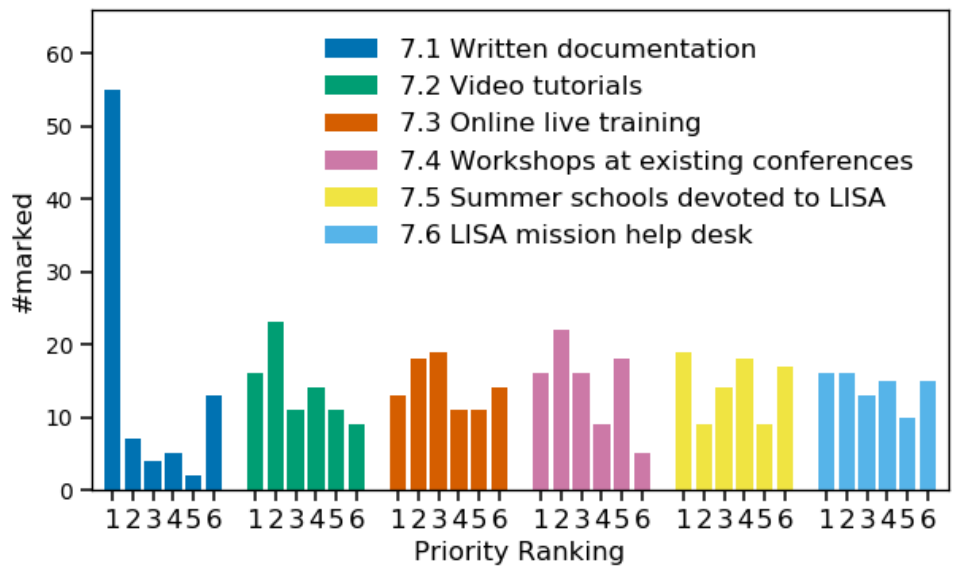

*Somewhat Familiar* or *Unfamiliar* Respondents:

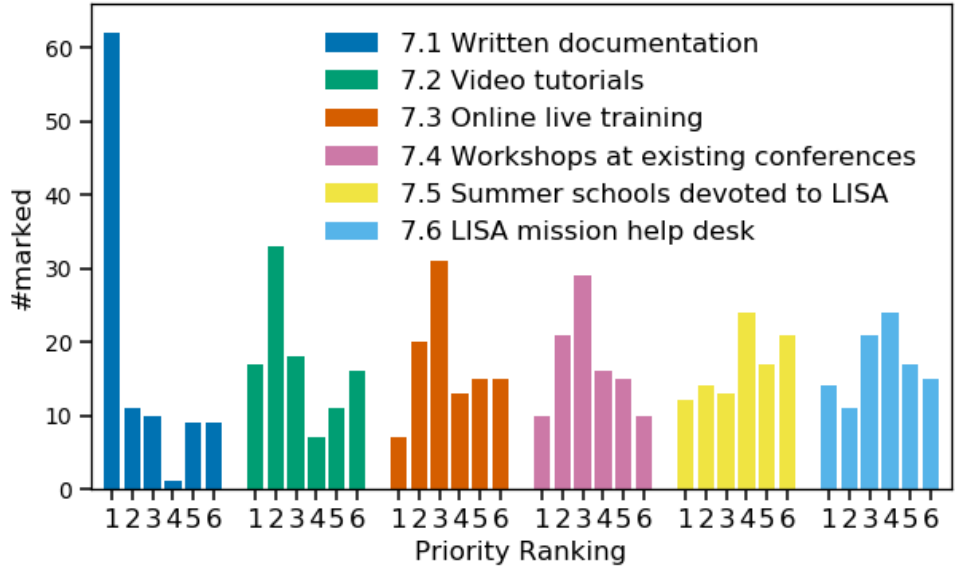

**8. When would you like to take LISA data-training?** (mark only one oval)

8.1 As soon as possible (using mock data).
8.2 Close to the launch date (early 2030's).
8.3 When the first LISA data release occurs (mid 2030's).
8.4 Not interested in LISA data-training.

### *Very Familiar* Respondents:

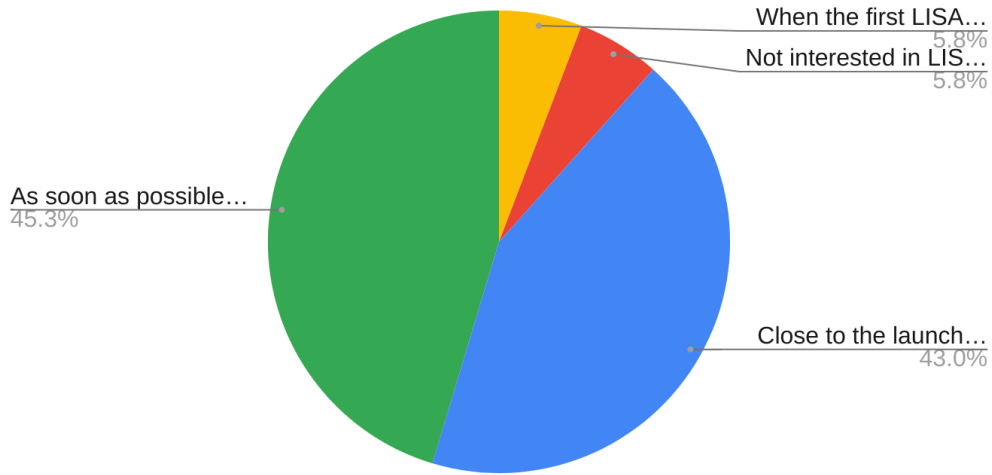

8. When would you like to take LISA data-training?
- When the first LISA… 5.8%
- Not interested in LIS… 5.8%
- Close to the launch… 43.0%
- As soon as possible… 45.3%

### *Somewhat Familiar* or *Unfamiliar* Respondents:

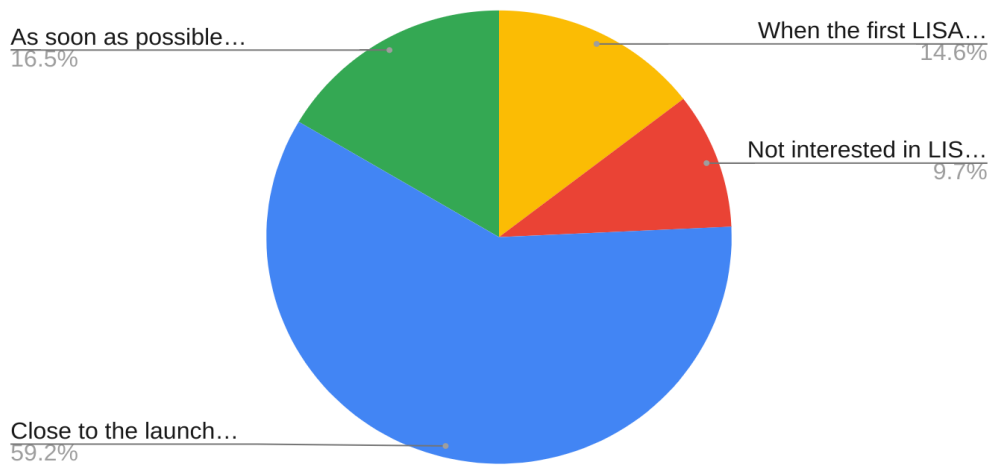

8. When would you like to take LISA data-training?
- As soon as possible… 16.5%
- When the first LISA… 14.6%
- Not interested in LIS… 9.7%
- Close to the launch… 59.2%

**9. How interested are you in the following initiatives?** (Please rank-order the options, where 1=most interested and 5=least interested.)

9.1 Graduate and undergraduate curriculum in gravitational wave astronomy.
9.2 Fellowships for research partnerships between the gravitational wave and astronomy communities.
9.3 Faculty, postdoc, and student internships in LISA research groups.
9.4 LISA consultants to provide tailored training and advice on data analysis.
9.5 Faculty cluster hires in gravitational wave astronomy.

*"Very Familiar" Respondents:*

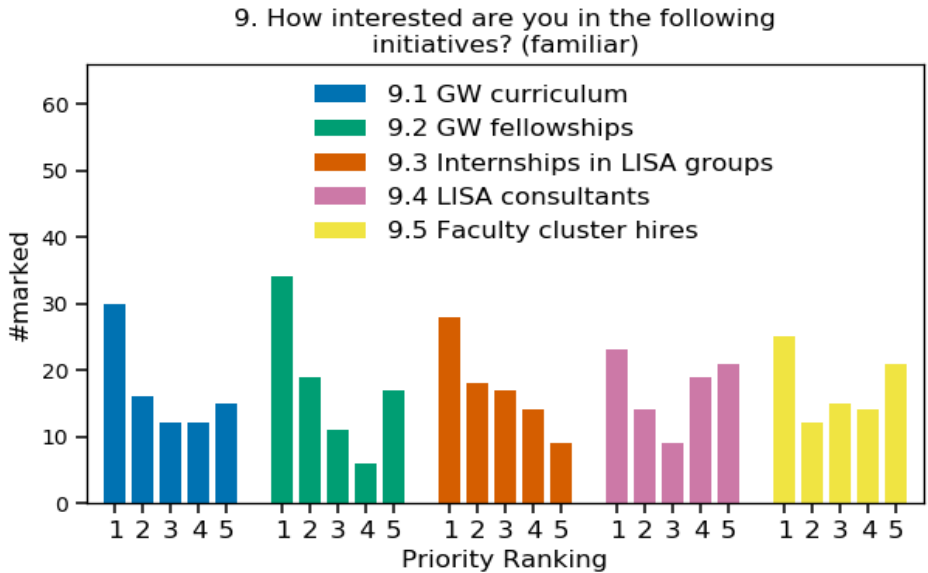

*"Somewhat Familiar" or "Unfamiliar" Respondents:*

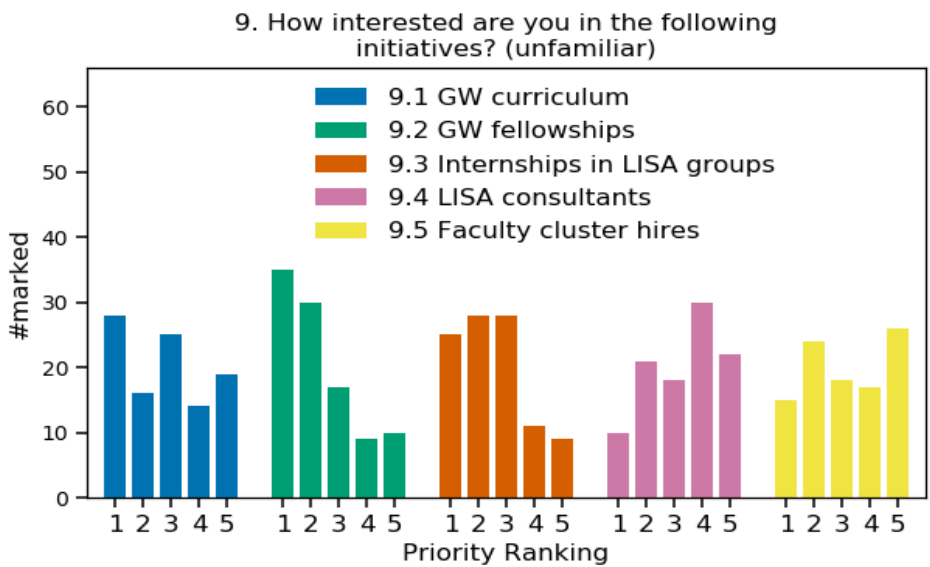

**10. Would you be willing to join a large collaboration to use the following LISA data?**
(Yes or No)

10.1 Alerts for follow-up observation.
10.2 Catalog of resolved sources, both continuous and transient.
10.3 High level data (e.g. all-sky strain maps) with Galactic foreground.
10.4 High level data with Galactic foreground removed, plus foreground model.
10.5 Frequently updated but less accurate 'quick look' data for monitoring potential sources.
10.6 Low level data with gaps, glitches, instrument noise, plus access to select engineering channels.

*Note: This is Question 3 asked a different way; the Yes/No responses are correlated with the interest-level reported in Question 3.*

### *Very Familiar* Respondents:

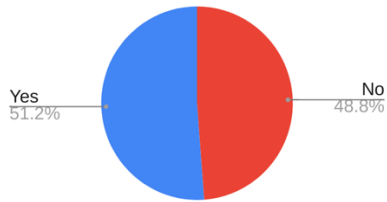
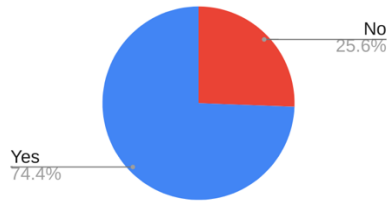
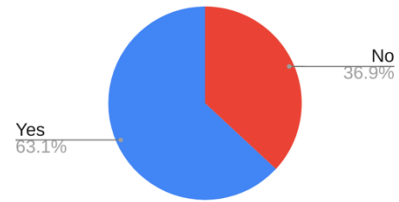
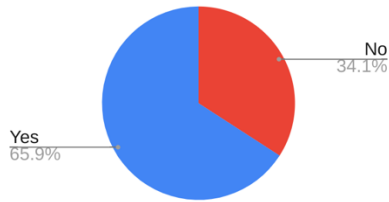
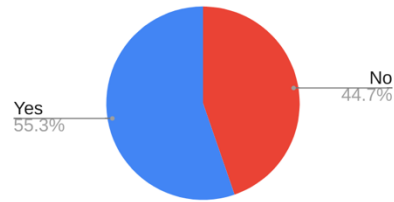
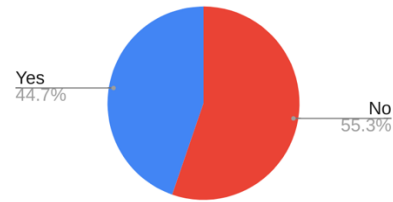

### *Somewhat Familiar* or *Unfamiliar* Respondents:

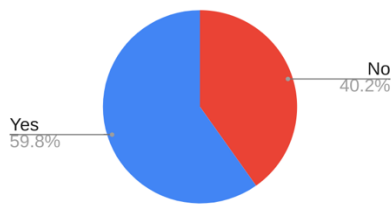
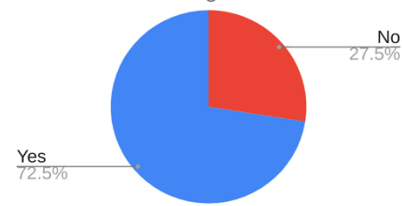
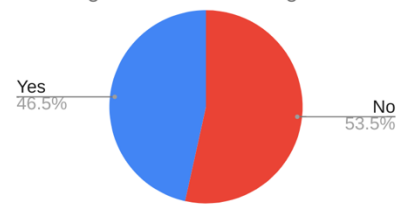
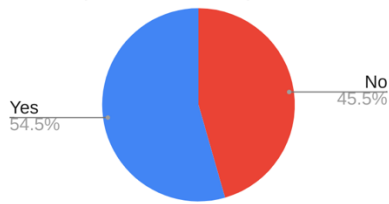
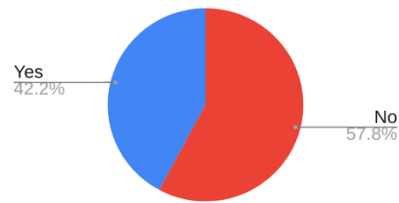
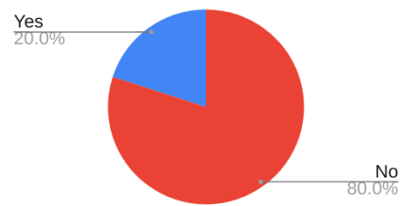

**11. What precursor science should be done ahead of launch (i.e. model black holes; survey potential host galaxies; observe Galactic binaries; explore mock data)?**

*Very Familiar* Respondents
- All of the above. (x2)
- Explore mock data. (x10)
- Model sources. (x7)
- Observe Galactic Binaries. (x13)
- Survey potential host galaxies. (x5)
- Astrophysical prediction of potential LISA impact beyond and conjunction with future. Virgo-Kagra-LIGO at 2030.
- Black hole in extended gravity models, alternatives for black holes.
- Calculation of 2nd order gravitational self-force.
- Catalog all nearby binaries, predict emission properties for mergers, perform detailed rate calculation for comparison with measured.
- Compact object merger models, mock data, survey of potential hosts.
- Data analysis methods as opposed to source modeling .
- Develop accurate waveform models, explore mock data.
- Develop techniques for subtraction and parameter estimation of confusion noise.
- Enhance awareness of the astronomy community.
- eRosita source follow-up.
- Explore mock LISA data, taking into account upcoming survey data like the Vera Rubin Observatory's LSST camera.
- Fractals in binaries.
- Further exploration of the LISA "global fit" problem through mock data analysis
- Gravitational wave source modelling; (mock) data analysis pipelines; modelling of astrophysical sources, their properties and event rates; modelling to explore fundamental physics questions; complementary astrophysical surveys.
- Ground-based GW observing to help inform and plan the LISA mission.
- Improve galaxy catalog.
- Instrument modeling.
- Merging double white dwarf discovery surveys - RV & transit method work (I may have been involved in such so a little bias may be there :).  Sources constran galactic foreground models, merger products/SNe rates, for initial LISA calibration, etc.
- Mock data should probably be priority #1, so people can start analysis very quickly post first-light.
- Model and find LISA sources from EM.
- Model SMBH mergers, EMRIs, Develop new data analysis techniques.
- Model the complete expected low frequency GW signal for f <= 1 Hz .
- Monitor Galactic binaries, find them and study them in detail in EM (masses, periods, sky position, Pdots), provide mock data based on real EM sources to understand how GW and EM can complement each other.
- No idea.
- Numerical modeling of SMBH binary evolution
- Optimizing source detection algorithms and testing with data challenges.

- Other estimates of expected black hole merger rates, from a variety of sources (e.g., galaxy mergers, simulations, AGN luminosity functions, etc.).
- Precise catalog of Galactic binary sources. Theoretical understanding of growth of massive black holes throughout cosmic time.
- Prepare supporting catalogs from current and near future surveys.
- Searching for which PTA sources will move into the Lisa band.
- Short cadence Galactic Plane data with LSST is vital to be properly prepared. Follow-up of supermassive BH mergers will likely require ngVLA, so that should be supported by the LISA team.
- SMBH/IMBH binary surveys; observation of stellar BHs around SMBH/IMBH (or likely environment for such); continuing galactic Re-usewhite dwarf binary survey.
- Source Modeling, EM Follow-up Testing (how rapid is the PE for each new source? what is the initial sky localization? how do the uncertainties in these values change with LISA observation time?), explore mock data (what amount of template fitting error is acceptable? Understand how PE errors effect the entire source catalog).
- Survey potential host galaxies of binary SMBHs; analyze mock LISA data and allow this to be done by interested parties to better understand/characterize the expected signatures of different source types (both continuous and transient) and develop pipelines for data analysis
- Time series analysis tools.
- WDB waveforms including tidal effects, etc.; gas drag models for black holes (esp multiband sources); survey potential host galaxies; explore mock data for SNR of various effects in backgrounds and foregrounds--what can you pull out?

### *Somewhat Familiar* or *Unfamiliar* Respondents

- All of the above. (x1)
- Explore mock data. (x10)
- Model black holes. (x8)
- Observe Galactic binaries. (x11)
- Survey potential host galaxies. (x11)
- All sky survey for galaxies, data modeling.
- Binary studies in nearby low-Z galaxies.
- Calculation of GW cosmological wave remnants from very early universe.
- Collapsar source models.
- Complete catalogues and characterisation of galactic binaries.
- Construction of the model library for binary black holes and black hole merger.
- Develop electromagnetic follow-up strategies.
- Estimates of binary SMBH incidence based on E/M studies of mergers /post-merger sources.
- Explore synergies with LSST and WFIRST which should identify variable AGN.
- Explore theoretical sources.
- Galactic modeling, binary evolution modeling, better understanding of precursors to GW systems.
- Gamma and xray searches. Broad-band gamma's with decent (1h) time resolution.
- Intensive long term study of variable galactic sources.
- Mock testing of a rapid response system for EM follow-up observations.
- Model compact object binaries from different formation channels.

- Model electromagnetic counterparts, e.g. to high-mass ratio mergers.  Find galactic binaries detectable with LISA.  Make predictions from different population models on the number, mass, etc. of binaries that LISA will detect.
- Model expected populations. (x2)
- Model gravitational waves without surface wave properties.
- Model mergers within galactic environments to determine features and/or existence of electromagnetic counterparts.
- Modeling of all of the most frequent and also the loudest sources.
- Models of LISA emission and simulated corresponding EM signals from sources that are within the detection thresholds.  Matching to existing galaxy survey data, and discussion of upcoming surveys and missions (e.g. LSST, WFIRST, future NASA/ESA flagships) that would overlap with LISA and would be capable of following up extragalactic signals.
- Modified gravity models, numerical simulations for source signals and analysis of signal propagation.
- My interest is infrared time-domain monitoring of host galaxies and follow-ups of triggered events.
- Numerical relativity modeling.
- Observe low-metallicity extragalactic binaries.
- Observe potentially overlooked sources of galactic foreground.
- Observe the nature of gravitational waves, predict future discoveries of this general relativity-based field after LISA's launcing.
- Pathfinding with Context Perspective in Unicode.
- Prediction on electromagnetic counterparts.
- Predictions of galactic sources from mock data and simulations.
- Rates of sources, and especially get an idea of the uncertainty on the distance, masses, chirp mass, mass ratio etc. for the sources.
- Simulations of close binary black hole evolution.
- Simulation of SMBH mergers.
- Solutions to astrophysical problems related to multi-messenger science using LISA (e.g., AGN variability).
- Standard siren mock analyses.
- Theoretical models about  black holes and explore mock data.
- Theoretical probability analysis of potential polarization signatures for various transient ssurces producing GWs.
- Theory and observations of merging super-massive black hole binaries.
- X-ray survey to establish baseline of activity.

**12. Which best describes your career stage in the field?** (mark only one oval)

12.1 Undergrad or before.
12.2 Graduate student.
12.3 Early career (0-7 yrs since degree).
12.4 Mid career (8-15 yrs since degree).
12.5 Late career (15+ yrs since degree).

### *Very Familiar* Respondents:

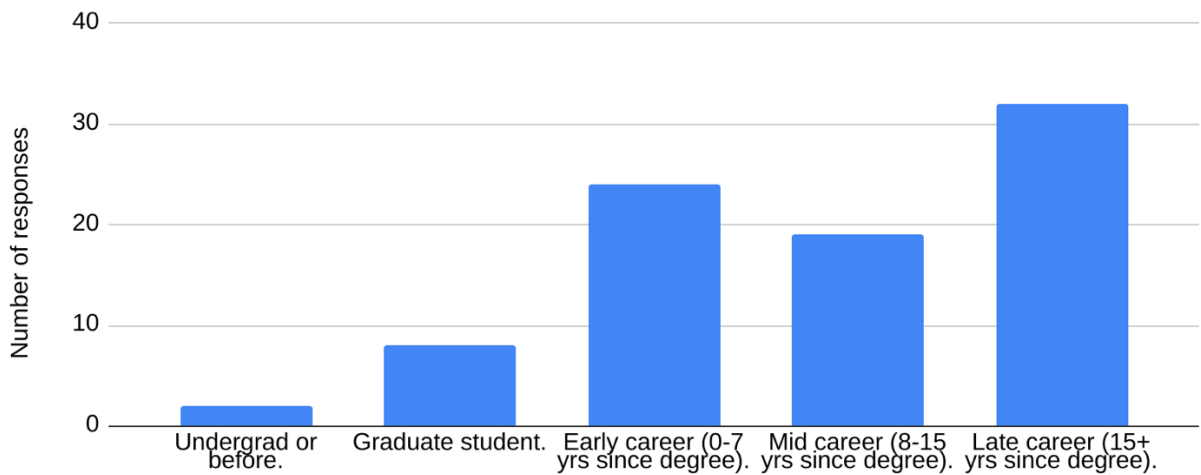

### *Somewhat Familiar* or *Unfamiliar* Respondents:

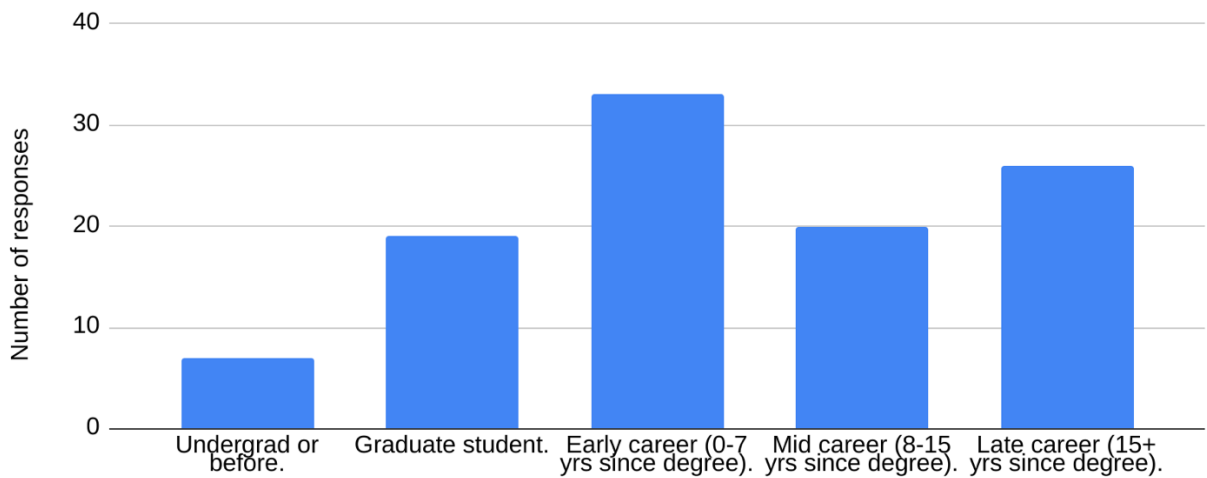

**13. How would you describe your research (e.g. theory; optical astronomy)?**

### *Very Familiar* Respondents

- Astronomical & Physical technology development
- Astrophysics
- Broad
- Compact objects theory and observations
- Computational, source modeling
- Data-analysis. (x2)
- Detector Characterization at for LIGO and Virgo. (x2)
- Dissertation topic is on the 12-minute eclipsing WD. Ultimate interest is theoretical gravity and exploration of the particle aspect of quantum theory.
- Extragalactic Astrophysics
- fundamental physics
- galactic astronomy, astroinformatics
- Gamma Ray Astronomy
- Gravitational wave astronomy. (x7)
- Gravitational wave data analysis. (x4)
- Gravitational wave sources theory, also laser and particle physics
- Gravitational wave theory (source modelling of black-hole binaries, tests of GR)
- Gravity and Cosmology Theory
- High energy astro theory
- History of physics & astronomy; scientometrics; reviews
- Infrared astronomy
- Instrument building and communication
- Instrumentation for Gravitational wave detection
- Instrumentation for ground based GWD. But these questions are answered in an outreach capacity.
- Interacting binary stars
- Laser interferometry
- LIGO data analysis
- Mainly optical astronomy, Gaia variable stars and astrometry, some radio interferometry
- Multi-messenger astronomy with VHE gamma-rays
- Multi-wavelength transient
- Multiwavelength astrophysics, theory and observation
- Non-linear phenomena in strong gravity
- Observational cosmology
- Observational high energy astrophysics. (x2)
- Observational multiwavelength
- Optical Astronomy (x3)
- Optical Astronomy, LIGO Data
- Optical exatragalactic observational
- Optical time-domain astronomy astronomy
- Previous researcher in Mock LISA data challenges; now retired from NASA

- PTA astronomy. (x2)
- Radio. (x2)
- Radio Astronomy / VLBI / Precision timing / Pulsar
- Radio astronomy, neutron stars, gravitational waves
- Staff LIGO in data visuakization and detector characterization
- Supernovae, high-energy astrophysics, nuclear
- Theoretical and experimental gravitational physics
- Theory. (x21)
- Theoretical galaxy formation
- Theoretical relativity
- Theory, computational and data analysis
- Theory, gw data analysis
- Theory, model testing, cosmology
- Theory, waveform modelling
- Theory, but currently only distantly related to LISA/GW astronomy
- Theory/computational.
- X-ray astronomy.
- X-ray observations of Galactic compact objects, including black hole binaries

### *Somewhat Familiar* or *Unfamiliar* Respondents

- AGN and quasar observations
- Applied science
- Astrochemistry
- Astrometry; Stellar Occultation
- Computational. (x3)
- Cosmology theory, data analysis
- Data-driven extragalactic astronomy
- Education
- Experiment; radio astronomy
- Gamma-ray astronomy. (x2)
- Gravitational microlensing; between theory and observation
- Gravitational waves data analysis, Cosmology
- GRB, Stellar evolution
- High energy astrophysics: both computational/theoretical plasma astrophysics and X-ray astronomy.
- Infrared astronomy. (x3)
- LIGO detector characterization
- Millimetre astronomy
- Multi-wavelength observational astronomy. (x3)
- Multi-wavelength observational studies of galaxy evolution including AGN feedback processes.
- Multifrequency and multimessenger studies of transients

- N-body simulations
- Neutrino particle astrophysics (x2)
- Numerical relativity
- Numerical simulations and optical astronomy observations / data analysis
- Observational Astronomer (x14)
- Optical / Infrared observational astronomy (x8)
- Pulsar astronomy
- Radio astronomy (x3)
- Scalable applied game theory
- Stellar populations
- Survey cosmology
- Theory (x11)
- Theory anchored in observations
- Theory and high-energy observation
- Theory with interest in possible quantum gravity signatures
- Theory, exoplanets
- Theory, modelling
- Theory; stellar astronomy
- Theory: gravitational waves, astro-statistics
- Transient observer
- UHE neutrino experiment
- X-ray astronomy (x5)
- X-ray/optical/NIR astronomer

**14. What is your institution/affiliation?**

*Very Familiar* **Respondents**

- Andrews University
- APC laboratory, University Paris-Diderot
- Berea College
- California State University Fullerton (x3)
- Caltech
- CEA Paris-Saclay, France
- CfA
- CIERA/ Northwestern University
- Cornell University (x2)
- CUNY-AMNH
- Department of Physics and Astronomy, Vanderbilt University, Nashville, TN
- Flatiron Institute
- Florida Institute of Technology
- Geneva Observatory
- Georgia Tech
- Grand Valley State University
- Harvard
- Higher Education
- Hillsdale
- IISER Kolkata, India (x2)
- INAF of Padova, Italy, and U Wisconsin, USA
- Johns Hopkins University Applied Physics Lab
- JPL/Caltech
- Kavli Institute for Theoretical Physics
- Long Island University
- Max Planck Institute for Gravitational Physics (AEI)
- MIT (x3)
- NASA GSFC (x2)
- National Centre for Nuclear Research, Poland
- Nikhef, Amsterdam
- None (post PhD graduation unemployment)
- Northwestern
- NSF's OIR Lab
- Observatoire de la Côte d'Azur (France)
- ORBIT
- Penn State
- Previously NASA Ames, now retired
- Rochester Institute of Technology (x2)
- Sapienza University of Rome
- Space Initiatives Inc
- Texas Tech University

- The Ohio State University
- U California Irvine, Queen Jadwiga Observatory
- UBC
- UCSB
- Univ. of Arkansas
- University of Notre D
- University of Birmingham (x3)
- University of Chicago (x3)
- University of Florida (x3)
- University of Illinois at Urbana-Champaign (USA)
- University of Illinois, University of Cambridge & QMUL
- University of Louisville
- University of Michigan (x2)
- University of Minnesota
- University of Missouri
- University of New Mexico
- University of São Paulo, Brazil
- University of Trieste
- University of Zurich
- Uppsala University
- UW
- UW-Milwaukee
- Vanderbilt University
- WVU
- Yale University

### Somewhat Familiar or Unfamiliar Respondents

- Ben Gurion University
- BR, Salvador / BA CityHall
- Brigham Young University
- California State University Stanislaus
- Caltech (x2)
- Cardiff University
- CCA, Flatiron Institute
- Center for Astrophysics | Harvard & Smithsonian (x2)
- Clemson
- CNRS
- Eastern New Mexico University
- ESO
- Florida State Astronomy
- Gamerjibe, Bagelbyte, Zeolite Studios
- Gemini Observatory
- Hanoi National University of Education, Hanoi, Vietnam
- Harvard

- Heidelberg university
- Hofstra University
- INAF-IRA Bologna
- INAF-OAS
- INAF/OAB
- Indiana University Kokomo
- JPL/NASA
- Karnataka Open University
- Max Planck Institute for Gravitational Physics
- McGill University
- Michigan State University
- Morehead State University
- MPE
- MPIA
- NASA
- NASA/GSFC (x3)
- National Observatory of Athens
- National Radio Astronomy Observatory
- Northwestern University and the Adler Planetarium
- NRAO
- NREL/LANL
- NSF OIR Lab
- Observatoire de Paris
- Ohio State University (x4)
- Pennsylvania State University
- Princeton University (x2)
- PSU
- SMA Negeri 1 Gombong
- Smithsonian (x2)
- Stanford University
- Stanford/SLAC
- Texas Tech University
- The Pennsylvania State University
- The University of Alabama
- The University of Texas at San Antonio
- Tsinghua University
- TTU
- UC Berkeley
- UC Irvine (x3)
- UMass Lowell
- United States Naval Observatoy (Flagstaff Station)
- University of Arizona (x5)
- University of Birmingham
- University of British Columbia/LIGO
- University of Colorado Boulder

- University of Connecticut
- University of Heidelberg
- University of Illinois at Urbana-Champaign
- University of Michigan (x4)
- University of North Dakota
- University of Toronto
- University of Wisconsin-Milwaukee
- Vanderbilt University (x6)
- Wayne State University
- West Virginia University
- ZAH, University of Heidelberg

**15. Please provide your email if you wish to receive LISA-related emails.**

*Very Familiar* **Respondents:** (69%) 60 of 87 provided email addresses.

*Somewhat Familiar* or *Unfamiliar* **Respondents:** (70%) 74 of 106 provided email addresses.

**16. Do you have any further thoughts regarding your potential needs to maximize the usefulness of LISA data?**

*Very Familiar* **Respondents**

- Clear synopses of where the data comes from on the spacecraft, as well as what the data means, especially for high-level data. Re-using current tools/portals to EM databases would make exploring the high-level data easier.
- Collaboration shouldn't be about data, but about tools and understanding.
- Connection of LISA alerts for Ground based GW Detection.
- Considering python seems to have significant staying power, it might be helpful to produce a library, or collaborate with astropy (the latter probably being easier), that would allow astronomers to do certain pieces of the analysis more easily.
- Don't make the same mistake as LIGO, release the data as quickly as possible so that the wider community can interface with it easily.
- Estimation of alert rates, time delays until emission of alerts and localisation uncertainties in 3D.
- How to couple the knowledge I have about LIGO data analysis to LISA analysis.
- I am a theorist watching these developments from outside but I think it's wonderful.
- I married into the field (widow of Joseph Weber) and remain interested in all aspects.
- I would like to attend an online workshop.
- In order for current graduate students to make contributions to LISA science, it is important that the pipeline for getting those graduate students into postdocs and ultimately faculty or permanent research positions is built now, so that researches with an extensive knowledge base can be prepared to make the best possible use of LISA data, and mentor their own graduate students and postdocs when LISA launches in the 2030s. Financial and community support for precursor science in the 2020s, including source modelling, direct LISA modelling, and electromagnetic surveys to find and study LISA's galactic sources ahead of time are vital to ensuring there will be an active and experienced pool of scientists to promote LISA's science output.
- Keep up the great work!
- Looking at the records of past facilities, the biggest impact is had when a large segment of the science community is involved; and the latter happens when high-level data products are provided, rather than requiring outsiders to acquire black belts in analysis (although of course a few would prefer the latter kind of access). Widespread uptake of open surveys, catalogs, *digested* information, available to the broadest community (and not just the constituent community) - this is what leads to high impact on science.
- Make the data as open as possible. produce collaboration-wide papers describing tools -- these can then be cited by all users of the data so that credit is given.

- Maximizing the usefulness of LISA data is good.

- Not yet, but we see where I am in 10 years from now …

- Open source and public data, limited data priority.

- Sustained moderate funding required.

- The broad astro community is going to need help with LISA data--there needs to be a data center that provides both training and live help, especially to get reliable parameter estimations with non-standard source models (due e.g. to gas effects for small mass BH).

- The more the merrier!

- Thorough understanding of the instrument and associated noise, glitches, calibration etc. is necessary to maximize the fidelity of science that can be done with LISA. This is on top of the more commonly considered activities like source modeling. Understanding the instrument is more in my area of research, hence my preference for low-level data.

- With the first data release maybe 15 years out (half a career!), a disconnect between the GW folks and those of us working in astronomy, and white dwarf science being a very marginal part of GW communities current young talent like myself is going elsewhere. The GW community wants to hire GW people, and the astronomy community wants to hire astronomers from grants that are not for GW science. Consider cross over programs between the fields to blend ideas and expertise, and try to fund multimessenger astrophysics. Time will tell if I return to the astronomy side of things when the LISA data finally begins to trickle down, but the most pressing way to maximize what I might do with LISA data is for jobs to open up for an optical astronomer on the GW side of things now.

- Yes: guidance for obtaining access to gravitational theory groups where I can work on my nascent theoretical concepts regarding quantum gravity.

- LISA will be transformative – in the science, in the data, and in the mission itself. With such a huge potential for a paradigm shift, it's crucial to try to futureproof the data analysis with extensive documentation, lots of data from intermediate steps in the analysis, and archived information all the way down to the nuts and bolts of the spacecraft to triple-check amazing claims (think the Opera experiment and tachyons that were really some fiber optic glitch in the clock). One aspect of futureproofing is getting the broadest community involved to look at the data and

### *Somewhat Familiar* or *Unfamiliar* Respondents

- Alert system to implement rapid follow-up.

- I think the LISA project should give better info, even accelerate our universe's thought, so it's important to always update the development of this project and gain all the possible resources from other affiliates that also involved in this program.

- It would be great to have a dedicated telescope with a high-throughput, low-moderate-resolution optical spectropolarimeter that can be used for rapid followup of LISA detections.

Such a resource will be able to follow the both the spectral and polarization op optical counterparts to LISA GW detection and provide unique insights into the physics of events that produce detectable GWs.

- It would be nice to know about relevant uncertainties for maximizing the efficiency of Fisher forecasting analyses for the near term future.

- LISA data could be made especially useful if LISA is launched to orbit successfully.  Don't spend too much money >3 years before launch on scientists re-doing simulations and trumpeting how great LISA will be. Get the launch done on time."

- Make it public and release associated quick look products in a rapid manner.

- maybe discussing what thresholds LISA will choose in the future and why, and especially considering burst and stationary sources.

- Nope, not yet. Need to learn more details.

- Public data is the best data.

- Question 6 didn't allow multiple selections at the same rank; I would rank the last two entries both at #2.

- Report positive discoveries.

- The LISA community needs to provide introductory material to LISA signal detection and data analysis for undergraduate and graduate students to connect to the astronomical and physics communities in the coming decade.

- Turn data into pictures on a simulation.

- Understanding more precise consequences which affected by g waves, such change in electric potential between plates placed along the direction of propagation of g wave.

- Unicode E0.0

- User friendly numerical GR tools would be interesting.

# APPENDIX C - LISA SCIENCE OBJECTIVES AND INVESTIGATIONS

The following list of LISA Science Objectives (SO) and Science Investigations (SI) was taken from Table 1 of the LISA Science Requirements Document (Iss. 1, 14 May 2018).

**SO 1 Study the formation and evolution of compact binary stars in the Milky Way Galaxy**

SI 1.1 Elucidate the formation and evolution of Galactic Binaries by measuring their period, spatial and mass distributions

SI 1.2 Enable joint gravitational and electromagnetic observations of galactic binaries (GBs) to study the interplay between gravitational radiation and tidal dissipation in interacting stellar systems

**SO 2 Trace the origin, growth and merger history of massive black holes across cosmic ages**

SI 2.1 Search for seed black holes at cosmic dawn

SI 2.2 Study the growth mechanism of MBHs before the epoch of reionization

SI 2.3 Observation of EM counterparts to unveil the astrophysical environment around merging binaries

SI 2.4 Test the existence of intermediate-mass black holes (IMBHs)

**SO 3 Probe the dynamics of dense nuclear clusters using extreme mass-ratio inspirals (EMRIs)**

SI 3.1 Study the immediate environment of Milky Way like massive black holes (MBHs) at low redshift

**SO 4 Understand the astrophysics of stellar origin black holes**

SI 4.1 Study the close environment of Stellar Origin Black Holes (SOBHs) by enabling multi-band and multi-messenger observations and multi-messenger observations at the time of coalescence

SI 4.2 Disentangle SOBHs binary formation channels

**SO 5 Explore the fundamental nature of gravity and black holes**

SI 5.1 Use ring-down characteristics observed in massive black hole binary (MBHB) coalescences to test whether the post-merger objects are the black holes predicted by General Theory of Relativity (GR)

SI 5.2 Use EMRIs to explore the multipolar structure of MBHs

SI 5.3 Testing for the presence of beyond-GR emission channels

SI 5.4 Test the propagation properties of gravitational waves (GWs)

SI 5.5 Test the presence of massive fields around massive black holes with masses larger than $10^3 \, M_\odot$

### SO 6 Probe the rate of expansion of the Universe

SI 6.1 Measure the dimensionless Hubble parameter by means of GW observations only

SI 6.2 Constrain cosmological parameters through joint GW and electro-magnetic (EM) observations

### SO7 Understand stochastic GW backgrounds and their implications for the early Universe and TeV-scale particle physics

SI 7.1 Characterise the astrophysical stochastic GW background

SI 7.2 Measure, or set upper limits on, the spectral shape of the cosmological stochastic GW background

### SO 8 Search for GW bursts and unforeseen sources

SI 8.1 Search for cusps and kinks of cosmic strings

SI 8.2 Search for unmodelled sources

# APPENDIX D - DETAILED LATENCY CONSIDERATIONS BY SCIENCE OBJECTIVE

**SO 1: Galactic Binaries:** Electromagnetic detection and characterization of LISA galactic binaries contributes to *SO 1: Study the formation and evolution of compact binary stars in the Milky Way Galaxy*. Galactic binaries in LISA's band can be identified electromagnetically through several channels at a variety of EM wavelengths. Most of these sources are steady-state, and are not merging; their parameters are not expected to significantly evolve over the span of a few years. Electromagnetic observations can then occur anytime after the discovery of the source. However, EM observations of galactic binaries can improve joint EM-GW constraints on many binary parameters. Consequently, the scientific impact of LISA will increase the earlier EM counterparts are found and characterized.

> *SI 1.1: Elucidate the formation and evolution of Galactic Binaries by measuring their period, spatial and mass distributions*: GW measurements of the period of a binary are a prerequisite to conclusively identifying any EM counterpart. After the period is obtained, joint EM-GW observations of a binary will drastically improve measurements of the distributions of binary parameters and spatial locations. Therefore, the LISA collaboration will benefit from incentivizing EM precursor searches for galactic binaries and facilitating follow up searches for previously undetected ones. Follow-up searches can only begin once EM observatories capable of searching for short period binaries are alerted to the existence of the source.

> *SI 1.2: Enable joint gravitational and electromagnetic observations of GBs to study the interplay between gravitational radiation and tidal dissipation in interacting stellar systems*: By construction, this science investigation requires collaboration with EM observatories. One of the most powerful ways EM observatories can improve tidal dissipation studies with LISA is by providing an improved measurement of the orbital period derivative, which requires repeated observations of a system over an extended period of time. A delay of more than a few months in the release of a LISA-detected binary and subsequent localization of the EM counterpart would substantially degrade the quality of the period derivative measurement available by the end of LISA's primary mission. While period derivative measurements could continue after LISA's mission ends, such studies would then be produced by researchers outside of the collaboration. The measurements also could not be correlated with contemporaneous LISA data, ultimately reducing LISA's scientific impact on the study of tidal dissipation.

**SO2: Massive Black Holes:** *Tracing the origin, growth and merger history of black holes across cosmic ages* requires characterizing the environments in which massive black holes (MBHs) form. The impact of latency on this objective depends upon the specific science initiative:

> *SI 2.2: Study the growth mechanism of MBHs from the epoch of the earliest quasars*: This science investigation does not rely on obtaining EM counterparts, and progress could be made by comparing EM and LISA populations in archival data. However, if the merging MBHs are associated with accretion disks, the merger may result in a quasar flare or the launch of a jet, which could produce bright EM transients, after which deeper follow up observations could identify the host galaxies. Observation of EM transients originating from an MBH merger represents an opportunity for groundbreaking multi-messenger discoveries with LISA observations. The considerations are similar to those discussed below for *SI 2.3*.

*SI 2.3: Observation of EM counterparts to unveil the astrophysical environment around merging binaries*: By construction, this science investigation requires collaboration with EM partners. For masses $10^4$-$10^6$ M$_\odot$, a confident detection signal-to-noise ratio may be reached <1 day prior to merger. Combined with potential delays of several hours for downlink and signal processing on the ground, and initial localization areas as large as 100 deg$^2$, targeted EM searches must begin within hours to have a good chance of observing a counterpart. Such a response would require the development of a rapid alert infrastructure. As discussed above, public alerts for events with poor localizations enable large numbers of observatories to coordinate, reducing the dependence of a detection on latitude and weather at a specific site. Regardless of whether alerts are public, alerts must contain the candidate sky location to be useful, which should be updated as often as practical. To reduce wasted community resources on very distant sources and encourage follow up of closer sources, it is also crucial that any alerts contain the estimated distance. Unlike current LIGO detections of stellar-mass black hole mergers, the possible range of component masses for MBH mergers is many orders of magnitude. The large variation in component mass likely leads to differences in the type of EM signal for which one should search, and therefore including component masses in alerts may be more important for LISA than it currently is for LIGO. EM observers also may value spin information in alerts, as MBHs with aligned spins may be more likely to have accretion disks, which could trigger EM signals.

**SO3: Extreme Mass Ratio Inspirals (EMRIs):** *Probe the dynamics of dense nuclear clusters using EMRIs.*

Extreme mass ratio inspirals (EMRIs) occur when a less massive stellar origin compact object (especially a stellar-mass black hole) merges with a massive black hole. Aside from providing a probe of the stellar environment near massive black holes, these events will provide some of the most rigorous tests of general relativity (SO5). The rate at which EMRIs occur is currently very poorly constrained, as it is a sensitive function of star formation history, cluster dynamics, and mass segregation. Nearby EMRIs will have high signal-to-noise ratio and will be very well localized by LISA alone, improving the chances for EM observatories to find counterparts. Additionally, the early stages of an extreme mass ratio inspiral of a SOBH into an MBH are unlikely to eject a large accretion disk completely. Therefore, EMRIs are excellent candidates to possess observable EM variability due to interactions with and lensing of pre-existing accretion disks. Identifying and characterizing the interaction of a natal environment with a specific LISA source is a rich scientific opportunity to improve scientific understanding of star formation, cluster dynamics, and gas cloud physics.

The specific waveform of an EMRI as detected by LISA is likely to be correlated with the temporal variation of any EM counterpart, because the phase of the GW waveform is related to the lighter black hole's orbital phase. Therefore, the EM signature may exhibit characteristic fluctuations at specific orbital phases. Thus, to optimize EM searches and confidently identify possible counterparts, EM observers would value alerts containing the orbital phase information measured by LISA. Such products would need to be updated frequently as parameter estimates improve.

The discovery of an EMRI counterpart is so significant, regardless of competitors, that most observatories interested in such sources would participate without an MOU granting exclusive data access. The search for EMRI counterparts is therefore unlikely to be a strong motivator for increased partner participation.

**SO4-5 -- Stellar Origin Black Holes:** Completion of *SO4: Understand the astrophysics of stellar origin black holes* and *SO5: Explore the fundamental nature of gravity and black holes* will require the LISA collaboration to interface with both EM observatories and ground-based GW detectors. Months prior to coalescence, LISA will be capable of warning other observatories of the coalescence time of an event like GW150914 (the first stellar mass black hole binary discovered by LIGO) to within 1 minute and with sufficient sky localization to enable EM counterpart searches.

Here again, the impact of latency depends upon the specific science investigation:

> *SI 4.1: Study the close environment of SOBHs by enabling multi-band and multi-messenger observations at the time of coalescence*: By construction, this science investigation requires cooperation with other GW and EM observatories. Localizations within a few deg$^2$ will enable many EM instruments to promptly and deeply image the entire volume in multiple bands before and after coalescence. The resulting differential photometry will be higher quality than is possible for a source alerted only after coalescence, increasing the probability of detecting a faint counterpart beyond what is possible with only ground-based GW observatories.
>
> Early warning from LISA about an impending SOBH merger event with sufficient lead time (i.e., weeks to months) would allow data processing and alert transmission to occur well before the event. This allows EM observatories to be pointed and ready at the time of coalescence, making it possible to observe prompt (<10 s delay) merger counterparts. Alerting EM observatories to the time of coalescence as early as possible will also give them time to coordinate counterpart searches to minimize disruption to observatory scheduling, increasing the number of observatories potentially willing to participate. An embargo requiring the administrative overhead and commitments of an explicit data-sharing agreement to access alerts would reduce the potential for cooperation on this science investigation with observatories that might not be able to contribute significantly to other LISA science cases.
>
> *SI 5.3 Testing for the presence of beyond-GR emission channels*: This science goal requires testing the accuracy of LISA's GR-predicted coalescence times for SOBHs with the actual coalescence time, as observed by ground-based GW detectors such as LIGO, Virgo, KAGRA, and future successors. Early warning of the merger timing allows ground-based GW observatories to take steps to ensure they are operating in observing mode during the merger, such as rescheduling maintenance, commissioning, and calibration activities. Additionally, it allows observatories to make extra efforts to reduce detector noise to the lowest possible level around the coalescence. For example, they can preemptively disable hardware injections, restrict facility access to reduce foot and vehicle traffic, ensure loud equipment is not running nearby, etc.

SOBHs can also test deviations from GR using *Black Hole Spectroscopy*. Like electromagnetic spectroscopy, GW spectroscopy involves observing targets with instruments designed for different frequencies. If ground-based GW detectors are alerted to the predicted ringdown frequency of a bright GW source with sufficient lead time, they may be able to maximize sensitivity to the ringdown phase of the predicted coalescence. Although these goals can be attained without EM follow up, providing advance sky localizations to EM observatories to enable counterpart detection would allow a comparison between the GW and EM chirp signals, and thereby, probe theories in which the propagation speeds for gravitons and photons differ. Any deviation from GR or insight into gravitons would be a foundational, paradigm-shifting discovery.

**SO6-7: Cosmology:** *Probe the expansion rate of the universe* and *SO7: Stochastic GW backgrounds & implications for early Universe and TeV-scale physics* require LISA to collaborate with both external EM observatories and theorists, during the data interpretation.

*SI 6.1*: *Measure the dimensionless Hubble parameter by means of GW observations alone:* This science investigation requires comparing well-localized (<1 deg$^2$) sources with GW measured distances to the redshifts of all known galaxies in the localization volume to make statistical measurements of the Hubble parameter. This technique is possible even in the absence of specific EM counterparts, and becomes more useful as LISA's localization volumes get smaller and contain fewer galaxies. Some regions on the sky will already have reasonably complete archival galaxy catalogs out to the targeted redshifts for this technique (z<1.5), while other regions will require targeted follow-up from EM survey instruments for adequate completeness. Regardless of archival observations, deep follow up observations of a small localization region with EM observatories will always be able to obtain deeper catalogs, reducing the contamination of expansion rate measurements by selection biases against faint, distant, and heavily dust-obscured sources.

Selection bias will be one of the dominant sources of systematic error for this measurement. Therefore, it is highly desirable for LISA localization regions to be provided to EM observatories, to enable deeper multi-wavelength follow-up, especially at sky locations not thoroughly covered in archival data. EM follow-up for this particular purpose is not especially time-sensitive. Such deep follow-up imaging is, however, time consuming. It may be desirable to incentivize such follow-up by including participating observatories on high-impact LISA-published Hubble constant papers regardless of whether they do so as a result of a data-sharing agreement or an open public alert.

*SI 6.2***:** *Constrain cosmological parameters through joint GW and EM observations:* LISA will probe the expansion of the Universe using GW sirens at a range of redshifts. The observational concerns are essentially the same as the concerns discussed above for MBHBs for *SI 2.3: Observation of EM counterparts to unveil the astrophysical environment around merging.* The presence of MBHBs enables the acquisition of especially important data from high redshifts that few other methods can reach. GW standard sirens are most effective at constraining expansion when a counterpart is detected in coordination with EM telescopes so that a redshift can be measured directly. In particular, for distant, high-redshift galaxies, the galaxy catalogs required for *SI 6.1* will be highly incomplete. Therefore, the number of possible host galaxies within the uncertainty volume will be large. It is currently unknown how frequently these sources will have detectable EM counterparts. Sending alerts to partner observatories as quickly as possible will maximize the chances of locating any counterpart.

Even a single "golden" measurement, with a well-determined luminosity distance and redshift>3 would provide significant cosmological information and constraints on the values of cosmological parameters. Given reasonable rate estimates, a proprietary period of no longer than a few months would be adequate to publish a well-populated Hubble diagram, to increase the likelihood that EM counterparts could be found efficiently in such a scenario. After that point, open data would enable the community to re-analyze and extend the Hubble diagram as more events become available. Unaffiliated theorists could replicate much of the expansion rate analysis without deep collaboration expertise.

*SI 7.2: Measure, or set upper limits on the amplitude and spectral shape of the cosmological stochastic GW background*: As soon as LISA enters science mode, it will be possible to set non-trivial limits on the amplitude of an underlying cosmological mHz stochastic gravitational-wave background, for the simple reason that there are no other GW observatories competing in this frequency band. For the same reason, measurements of a stochastic GW background do not require any time-sensitive EM follow up. Therefore, the only way in which latency affects this objective is to provide a delay in external verification of results by the wider community.

.

**In the event only limits are obtained:** Roughly speaking, a good limit on a stochastic background down to 0.1 mHz can be placed within the first three months of LISA observing, to allow adequate sampling of the lowest frequency. The next significant milestone is after the first year of observing, upon sampling the entire sky, at which point the highly anisotropic astrophysical stochastic foreground can be better distinguished from a nearly isotropic cosmological background. To effectively search for signals the collaboration analysis has missed, external groups would require public releases of low-level spacecraft data, including raw phasemeter readouts. These scenarios would be affected by a prior detection of primordial GWs by other observatories (ground-based, including pulsar timing) as well as by CMB experiments.

**In the event a stochastic background is detected:** Signatures of TeV-scale events such as the electroweak phase transition will be extremely challenging to extract from the LISA data, but would have enormous payoff. Such an extraction will require substantial collaborative effort between the developers of instrument models and data analysis pipelines, theorists working on the contributions to the stochastic background from astrophysical sources, and early-universe theorists during the data interpretation. A high-confidence claim of a cosmological stochastic background requires access to LISA data at every available level, from raw phasemeter readouts and instrument telemetry to the most complete available known source catalogs. A claim of a cosmological stochastic background would be extraordinary, and no other currently-planned instrument could independently verify LISA's results. To achieve the highest level of scientific integrity and independent verifiability, all supporting data products would need to be made publicly available no later than the publication of any claimed cosmological stochastic background. Therefore, a responsible disclosure of a cosmological stochastic background is only possible after other proprietary periods for the periods of data analyzed have ended. For such an analysis to occur within the collaboration in the presence of any restrictive proprietary periods, the consortium will need to provide sufficient support resources to maintain a substantial theoretical presence within the consortium. If the proprietary period on such an important result were excessively long, there would be some risk of workforce trauma for collaboration theorists forced to sit on a potentially career-altering result for an extended time. However, provided information flow within the collaboration is unrestricted, and there is sufficient support for a theoretical community within the collaboration, this science case would suffer relatively little from a one or two year proprietary period.

**SO 8: Unforeseen sources:** *SO 8: Search for GW bursts and unforeseen sources* specifically targets pushing forward the discovery potential for gravitational-wave astronomy. Unmodeled sources will not have distances or parameter estimates available from automatic LISA pipelines, although sky localizations may be available. It is impossible to predict what kind of EM, neutrino, or astroparticle counterpart to such a source might be expected, or what observatories would be best capable of performing follow up.

For unmodeled events, there is an elevated risk that the instruments best able to detect counterparts would not have anticipated the need to sign an MOU with LISA. If counterparts to unmodeled events are missed due to the lack of an MOU with a capable partner, LISA could lose the opportunity to make once-in-a-lifetime revolutionary discoveries. Therefore, it is highly desirable for unmodeled events to be publicly alerted as quickly as possible.

Alerts for unmodeled sources would need to include somewhat different information than a typical alert to guide observers. This includes, but is not limited to, the sky localization, peak frequency, signal-to-noise, the start time and length of an event, the width in frequency space, and whether the event exhibits a chirp. Such an event will later require extensive data vetting and characterization to confirm the event and constrain possible progenitors, but any latency caused by hesitation to alert on an unknown event could result in missing a once-in-a-lifetime chance to obtain an EM or multimessenger counterpart. For these reasons, public alerts for unmodeled sources have the potential to drastically increase the scientific impact of the mission.

# APPENDIX E: DATA PRODUCTS

We expand on the data types produced at each level of processing in the following table. Note that this is only speculation, and not part of project planning at this time. Colored rows are nominally considered part of the AUX data product category for the purposes of this report.

|     | Data Product |
| --- | --- |
| PF | Performance models |
|     | Verification data |
|     | Calibration data |
|     | Performance test data |
|     | Spacecraft characterization data |
|     | Payload simulators |
|     | Spacecraft simulators |
|     | Constellation simulators |
|     | Control system models |
|     | Structural/Thermal/Optical/Performance models |
| ATM | Telemetry packets |
|     | Tracking data |
|     | Ground station data |
| L0 | Science data |
|     | Payload housekeeping data |
|     | Platform housekeeping data |
|     | Data quality metrics |
| L1 | TDI data streams |
|     | S/C ephemerides |
|     | Timing information |
|     | Auxiliary pathlength information |
|     | Noise characterization and stationarity metrics |
|     | Data quality metrics |
| L2 | Alerts |
|     | Global fit solutions – resolved sources |
|     | Global fit solutions – backgrounds |
|     | Global fit solutions -- residuals |

|  | Source identification and parameter estimates |
|  | Instrument noise models |
|  | Instrument response model |
|  | Data calibration models |
|  | Data correction models |
|  | Astrophysical models |
|  | Source waveforms |
|  | Supporting data from other observations |
|  | Alert Residuals |
|  | Data quality metrics (gaps, glitches, vetoes) |
|  | Algorithms |
|  | Source code |
|  | Processing history |
|  | Data products from intermediate iterations |
| L3 | Catalog of discrete sources |
|  | Catalog of background |
|  | Residuals |
|  | Instrument noise |
|  | Source models |
|  | Astrophysical models (galaxy, evolution, etc) |
|  | Source type statistics |
|  | Data quality metrics |